\documentclass[useAMS,usenatbib,fleqn]{mnras}
\usepackage{soul}

\usepackage{bm}
\usepackage[T1]{fontenc}
\usepackage[utf8]{inputenc}
\usepackage{multicol}
\usepackage{times,amsmath,amssymb,longtable,breqn}
\usepackage[varg]{txfonts}
\usepackage{grffile} 
\usepackage{graphicx}
\usepackage{color}
\usepackage[dvipsnames,table]{xcolor}
\usepackage{hyperref}
\usepackage{comment}
\usepackage{multirow}
\usepackage{caption}

\usepackage{enumitem}
\setlist[itemize]{leftmargin=6mm}
\setlist[enumerate]{leftmargin=6mm}

\usepackage{ulem}
\newcommand\al{\alpha}
\newcommand\gm{\gamma}

\newcommand\rs[1]{_\mathrm{#1}}

\newcommand\Esn{E\rs{sn}}
\newcommand\rhoism{\rho\rs{0}}
\newcommand\Mej{M\rs{ej}}
\newcommand\Mswept{M\rs{swept}}

\newcommand\Rshell{R\rs{sh}}
\newcommand\Mshell{M\rs{sh}}

\newcommand\Rch{R\rs{ch}}
\newcommand\Rbary{R\rs{bary}}
\newcommand\Rpwn{R\rs{pwn}}
\newcommand\Rrs{R\rs{RS}}

\newcommand\tch{t\rs{ch}}
\newcommand\tage{t\rs{age}}
\newcommand\Lch{L\rs{ch}}

\newcommand\tauz{\tau\rs{0}}
\newcommand\tbegrev{t\rs{beg,rev}}

\newcommand\Qinj{Q\rs{inj}}

\newcommand\gmbreak{\gm\rs{b}}
\newcommand\alL{\al\rs{l}}
\newcommand\alH{\al\rs{h}}

\begin{document}
\label{firstpage}
\pagerange{\pageref{firstpage}--\pageref{lastpage}}

\title[Reverberation of PWNe (III)]{Reverberation of pulsar wind nebulae (III):\\
%A self-consistent long term radiative evolution}
Modelling of the plasma interface empowering a long term radiative evolution}

\author[Bandiera et al.]
{R. Bandiera$^{1}$, N. Bucciantini$^{1,2,3 \,{\color{blue}\star}}$, 
B. Olmi$^{1,4}$\thanks{E-mail: barbara.olmi@inaf.it; niccolo.bucciantini@inaf.it} and D. F. Torres$^{5,6,7}$
\thanks{All authors have contributed equally to this work.}  \\
$^{1}$ INAF - Osservatorio Astrofisico di Arcetri, Largo E. Fermi 5, I-50125 Firenze, Italy \\
$^{2}$ Dipartimento di Fisica e Astronomia, Universit\`a degli Studi di Firenze, Via G. Sansone 1, I-50019 Sesto F. no (Firenze), Italy \\
$^{3}$ INFN - Sezione di Firenze, Via G. Sansone 1, I-50019 Sesto F. no (Firenze), Italy \\
$^{4}$ INAF - Osservatorio Astronomico di Palermo, Piazza del Parlamento 1, I-90134 Palermo, Italy
$^{5}$ Institute of Space Sciences (ICE, CSIC), Campus UAB, Carrer de Can Magrans s/n, 08193 Barcelona, Spain \\
$^{6}$ Institut d'Estudis Espacials de Catalunya (IEEC), Gran Capit\`a 2-4, 08034 Barcelona, Spain \\
$^{7}$ Instituci\'o Catalana de Recerca i Estudis Avan\c cats (ICREA), 08010 Barcelona, Spain \\}

\date{}
\maketitle

\pubyear{2023}

\begin{abstract}
%%%%%%% 242 parole
The vast majority of Pulsar Wind Nebulae (PWNe) present in the Galaxy is formed by middle-aged systems characterized by a strong interaction of the PWN itself with the supernova remnant (SNR). Unfortunately, modelling these systems can be quite complex and numerically expensive, due to the non-linearity of the PWN-SNR evolution even in the simple 1D / one-zone case when the reverse shock of the SNR reaches the PWN, and the two begin to interact (and reverberation starts).\\
Here we introduce a new numerical technique that couples the numerical efficiency of the one-zone thin shell approach with the reliability of a full ``lagrangian'' evolution, able to correctly reproduce the PWN-SNR interaction during the reverberation and to consistently evolve the particle spectrum beyond. 
Based on our previous findings, we show that our novel strategy resolves many of the uncertainties present in previous approaches, as the arbitrariness in the SNR structure, and ensure a robust evolution, compatible with results that can be obtained with more complex 1D dynamical approaches. 
Our approach enable us for the first time to provide reliable spectral models of the later compression phases in the evolution of PWNe.
While in general we found that the compression is less extreme than that obtained without such detailed dynamical considerations, 
leading to the formation of less structured spectral energy distributions, we still find that a non negligible fraction of PWNe might experience a super-efficient phase, with the optical and/or X-ray luminosity exceeding the spin-down one.
%Such super-efficiency might show up, depending of the PWN-SNR system, in the optical band rather than in X-rays.
\end{abstract}

\begin{keywords}
radiation mechanisms: non-thermal -- pulsar: general -- method: numerical -- ISM: supernova remnants  
\end{keywords}

%%%%%%%%%%%%%%%%%%%%%%%%%%%%%%%%%%%%%%%%%%%%%%%%%%
\section{Introduction}
\label{sec:intro}
%%%%%%%%%%%%%%%%%%%%%%%%%%%%%%%%%%%%%%%%%%%%%%%%%%

Pulsars (PSRs) are rapidly rotating and strongly magnetized neutron stars (NSs). 
The combination of rotation and magnetic field induces electric fields strong enough to pull particles out of the NS surface, to accelerate them to ultra-relativistic speeds, and ultimately to drive a pair-production cascade that results in the formation of a relativistic pair-plasma wind \citep{Michel:1973,Timokhin:2006}. 
The interaction of this wind with the environment (the parent SNR in the case of young PSRs), leads to the formation of relativistic wind bubbles, known as pulsar wind nebulae (PWNe) where the accelerated particles shine through non-thermal synchrotron and inverse Compton (IC) emission, with a very broad band spectrum extending from radio to TeV and PeV energies \citep{Gaensler_Slane06a, Slane:2017, Olmi_Bucciantini:2023}.

PWNe thus constitute an unique laboratory where high-energy astrophysical processes, non-thermal emission and relativistic plasma physics can be studied in great detail. 
For this same reason a plethora of different approaches have been put forward throughout the years, ranging from analytical to numerical one-zone models \citep{Pacini_Salvati:1973, Reynolds:1984,Gelfand_Slane+09a,Bucciantini_Arons+11a,Torres:2014}, from 1D hydrodynamics analytical \citep{Rees:1974,Kennel_Coroniti84a, Kennel_Coroniti84b,Emmering_Chevalier:1987} to 1D relativistic MHD numerical \citep{van-der-Swaluw:2001,van-der-Swaluw:2004,Bucciantini:2003,de-Jager:2008,Bandiera:2023}, from 2D MHD \citep{Del-Zanna:2004,Komissarov:2004,Bogovalov:2005,Del-Zanna:2006} to the more recent 3D relativistic MHD  \citep{Porth:2014a,Olmi:2016,Barkov:2019,Olmi_Bucciantini:2019}. 
For an exhaustive review on the various methods see \citet{Olmi_Bucciantini:2023}.

The fact that different approaches keep being developed and used, is a clear testimony to the fact that in the study of PWNe there is still no such thing as an \textit{all-purpose optimal strategy}, but that, depending on the issue at stake, different approaches might prove more suitable. 
There is now a general consensus that at least 2D, but more appropriately 3D, relativistic MHD is required in order to investigate the detail of the dynamics in the immediate post  termination shock region, and how this leads to the observed high energy X-ray jet-torus morphology of many systems \citep{Weisskopf:2000,Gaensler:2002,Lu:2002,Romani_Ng:2003,Camilo:2004, Slane:2004,Romani:2005}. 
However, this approach is so numerically expensive (millions of CPU hours to simulate just a few hundreds of years of evolution) that it is at the moment impossible to use it to model the long term %spectral 
evolution of even a single PWN. 
Moreover, none of these multi-dimensional HD or MHD models consistently includes the treatment of radiative losses and the dynamical spectral evolution of the source.
Indeed studies aimed at investigating the long term spectral evolution of many systems,  or that requiring the computation of several hundreds of models, as the case of population studies, as  in \citealt{Fiori:2021}, or multiple spectral fitting, as  in \citealt{Torres:2014}, are %still 
carried out using the so called one-zone models, where the PWN is treated as a uniform bubble, without internal structure, expanding within its parent SNR and subject to adiabatic and radiative losses.

All existing one-zone models adopt the same approximation, known as the ``thin-shell approximation'' \citep[see e.g.][]{Reynolds:1984, Bucciantini:2004a, Gelfand_Slane+09a, Martin:2016, Torres:2017}. 
As the PWN expands inside the parent SNR, it piles up matter into a shell (the swept-up shell). 
In the thin-shell approximation, such shell is assumed to be infinitely thin, and thus to perfectly trace the radius of the PWN. 
The shell evolution (a proxy for the PWN) can be obtained by solving the momentum conservation for the shell, which provides the acceleration (or deceleration) of the shell itself, due to the combined action of the PWN pressure from the inside, which acts as a piston pushing out, and the confinement of the outer SNR material.
The complete set of equations and boundary conditions can be found e.g. in \citealt[][Paper I hereafter]{Bandiera:2020}: a summary of them will be presented in the following section.
The thin-shell approximation has proven to be highly reliable in the study of young systems, where the PWN evolves within the cold self-similar expanding SNR ejecta, the so called free-expansion phase \citep{Gaensler_Slane06a}. 
Indeed it maps very well the 1D dynamics of young PWNe, as found by numerical simulations \citep{van-der-Swaluw:2001,Bucciantini:2003}.
The reason is twofold: in this phase the thickness of the shell is indeed much smaller than its radius (hence the correctness of the approximation), but even more important is the fact that the properties of the surrounding SNR ejecta are well defined  (namely they are cold and the expansion is self-similar, \citealt{Chevalier:1982}). 
This is quite important because all existing one-zone thin-shell models prescribe the outer structure of the SNR with simple analytical recipes.

Unfortunately, while the use of simplified recipes for the SNR is very robust in the early free-expansion phase, for the later reverberation phase, when the PWN interacts directly with the SNR shell, there is no consensus on how to model the confining environment, and this arbitrariness reflects in their increasing unreliability as the evolution proceeds beyond the free-expansion phase.
%there is absolutely no consensus on how to model the confining environment, with different groups opting for different recipes, often leading to substantially different results. \dft{this phrase is strong and I suggest to make it more precised. There not so many papers studying reverberation and in fact, from what I know more or less all of these studies are similar, e.g., \cite{Gelfand_Slane+09a, Vorster:2013, Torres_Lin:2018}} \nb{ [NB: except that in our Bucciantini et al 2011 we took a quite different approach to the reverberation phase, using a pressure rescaled according to numerical simulations for the SNR, halting mass accumulation in the shell, and truncating the compression at Sedov Equilibrium, without further bouncing, so we got a completely different evolution]} \dft{, as for results, it is just that some papers focused on one or the other thing, but underlying eqs. and approaches are about the same. I suggest to say:} {\bf "Unfortunately, while the use of simplified recipes for the SNR is very robust in the early free-expansion phase, for the later reverberation phase, when the PWN interacts directly with the SNR shell they become increasingly unreliable, and totally unreliable if one is interested in older systems well beyond this phase. }
%
%This makes results of one-zone thin-shell models totally unreliable if one is interested in old systems. 
%
Unfortunately, full 1D-HD models, able of capturing the full dynamics of the PWN-SNR system, are still too computationally expensive for population synthesis or multi-wavelength broad band spectral fittings.
But, at the same time, the main problem with one-zone thin-shell models  is that the internal dynamics in the SNR shell, and the feedback the PWN exerts on it upon interaction, are too complex for any simple recipe to fully capture them.

Recently \citet[][Paper II hereafter]{Bandiera:2023} have presented  non-radiative results based on one-zone models that however got rid of the thin-shell approximation, and solved for the correct structure of both the swept-up shell and the confining SNR using a full lagrangian approach. 
This allows for the one-zone evolution of the PWN  to be computed taking into account the correct SNR structure. 
Despite solving completely the ambiguity in the late evolution of all previous models, this approach is still too computationally expensive (it takes about a day on a single CPU to compute the full long term evolution of a single system). 
While this is undoubtedly the optimal avenue if one is interested in modelling just a few systems, it is not if one needs to compute hundreds or thousands of models, do fittings, or population analysis.
The major issue with this full lagrangian approach is that matter in the swept-up shell undergoes a strong compression, especially in the early phase of evolution, and this imposes a heavy computational burden to model those evolutionary phases. 
Interestingly, this is the same phase where the thin-shell approach has proved both robust and reliable. 

Driven by these considerations, we are presenting here a strategy for PWNe modelling that combines the best of both worlds: 
the efficiency and speed of the thin-shell approach and the reliability of the full lagrangian evolution. 
This is done by combining a thin-shell model for the PWN and swept-up shell \citep{Martin:2016,Martin_Torres:2022}, with a lagrangian model for the surrounding SNR \citep{Bandiera:2021}. 
Our approach is not as fast as a simplified one-zone thin-shell model (where a full PWN evolution can be run in a matter of few seconds), but fast enough (a full evolution can be run in just a few  minutes) to perform reliable population studies with multi-wavelengths spectral information and fittings.
We will show, by comparing with a full lagrangian approach, that this strategy is able to correctly reproduce the entire evolution of these systems from the very early free-expansion phase, to the late reverberation, capturing with accuracy the dynamics of the complex PWN-SNR interaction.

%%%%%%%%%%%%%%%%%%%%%%%%%%%%%%%%%%%%%%%%%%%%%%%%%%
\section{Thin-Shell-Lagrangian Code: TIDE+L}
\label{sec:tidevxx}
%%%%%%%%%%%%%%%%%%%%%%%%%%%%%%%%%%%%%%%%%%%%%%%%%%%%

Here we briefly illustrate the new computational strategy and how the thin-shell models are
%is 
coupled with a full lagrangian evolution of the SNR.

The evolution of the PWN can be divided into two phases: the early free-expansion and the late reverberation. 
The time at which the latter  begins and the former ends, $t_{\rm beg rev}$,  
occurs when the PWN swept-up shell (whose radius, $R_{\rm sh}$, is a proxy for the PWN radius $R_{\rm pwn}$ in the thin-shell approximation) equals the radius of the SNR reverse shock, $\Rrs$.

During the early free-expansion phase the swept-up shell evolves according to the standard thin-shell equations. 
One of those describes the conservation of mass:
\begin{eqnarray}
\Mshell(t) = 4\pi\int_0^{R_{\rm sh}(t)} \rho_{\rm ej}(t,r)r^2dr = \Mej\left(\frac{R_{\rm sh}(t)}{V_{\rm ej}t }\right)^3
\end{eqnarray}
where $\Mshell(t)$ is the mass in the shell at time $t$, $\rho_{\rm ej}(t,r)$ is the density distribution in the ejecta, and the last equality holds for ejecta with a uniform density profile.
In the same limit, the ejecta maximum velocity is given as a function of the ejecta mass $\Mej$ and the supernova energy $\Esn$ as: $V_{\rm ej}^2 =10\Esn/(3 \Mej) $.

Another equation describes the momentum conservation, 
under the mutual action of the PWN pressure $P_{\rm pwn}(t)$ from the inside and the confining ejecta from the outside:
%\begin{eqnarray}
%\frac{{\rm d}}{{\rm d}t}( \Mshell(t) \dot{R}_{\rm sh}(t) )= 4\pi \left[P_{\rm pwn}(t) -\rho_{\rm ej}(t,r)(\dot{R}_{\rm sh}(t) - R_{\rm sh}(t)/t  )\right]
%\end{eqnarray}
%\
\begin{equation}
{\color{black}
\frac{{\rm d} \left( \Mshell(t) \dot{R}_{\rm sh}(t) \right)}{{\rm d}t} =  4\pi P_{\rm pwn}(t)\,{R}_{\rm sh}^2(t) + \frac{\rm d \Mshell(t)}{\rm d t} \frac{R_{\rm sh}(t)}{t}}  \,,
%\frac{{\rm d} \left( \Mshell(t) \dot{R}_{\rm sh}(t) \right)}{{\rm d}t} \!\! = \! 4\pi R^2_{\rm sh}(t) \left[P_{\rm pwn}(t) - \! \rho_{\rm ej}(t,r)\left(\dot{R}_{\rm sh}(t) \! - \! \frac{R_{\rm sh}(t)}{t}  \right)^2\right],
\end{equation}
where $\dot{R}_{\rm sh}(t)$ is the expansion speed of the shell. 
The pressure in the PWN is computed taking into account the energy injection from the pulsar, as well as the adiabatic, radiation and diffusion losses as:
\begin{equation}
\frac{{\rm d}}{{\rm d}t}\left(4\pi R_{\rm sh}(t)^3 P_{\rm pwn}(t)\right) = L_{\rm psr}(t) - 4\pi  R_{\rm sh}(t)^2 \dot{R}_{\rm sh}(t)P_{\rm pwn}(t) - J(t)
\end{equation}
where $L_{\rm psr}(t)$ is the PSR spin down luminosity, and $J(t)$ factors in radiation and diffusion losses.

This set of equations is solved until $R_{\rm sh}(t) = \Rrs(t)$, where $\Rrs(t)$ is given, in the case of uniform ejecta,  by Eq.~22 in Paper II, which is itself based on a fitting of the numerical solution for the full SNR evolution (a more general formula for $\Rrs$ can be instead found in \citealt{Bandiera:2021}).

The same numerical solution for the full SNR evolution, not only provides the location of the reverse shock, but also the full structure (density, pressure, velocity) of the SNR, from $\Rrs(t)$ outward. 
This structure is the one used to initialize a lagrangian code for the evolution of the matter located outside $R_{\rm sh}(t)$. 
Moreover, at this same time,  we have the size $R_{\rm pwn}(t) = R_{\rm sh}(t)$, the expansion rate $\dot{R}_{\rm sh}(t)$,  the mass of the shell $\Mshell(t)$, as well as the interior pressure $P_{\rm pwn}(t)$. 
These are used to set the initial inner boundary condition of the lagrangian code, that evolves the matter outside, in the SNR.

For the reader convenience, we briefly summarize here the equations that are solved \citep{Mezzacappa_Bruenn+93a, Bandiera:2023}. 
The time evolution of the velocity ($v$) and radius ($r$) of the interfaces {\footnotesize{$i+1/2$}} between the shell {\footnotesize{$i$}} and the shell {\footnotesize{$i+1$}}  is given by:
\begin{align}
 H_{i}^n    \quad           =& \; \eta_{\rm vnr}\, \rho_i^n( v_{i+1/2}^{n}-  v_{i-1/2}^{n})^2 \, \Theta[v_{i-1/2}^{n} -v_{i+1/2}^{n}]\,,\\
  a_{i+1/2}^{n}     =& \;  [(r_{i+1/2}^n)^2(p^n_{i+1}-p^n_{i}) - (r_{i+1}^n)^2H_{i+1}^n + (r_{i}^n)^2H_{i}^n]/\Delta m_{i+1/2}\,,\! \\
  v_{i+1/2}^{n+1} = & \;  v_{i+1/2}^{n} - 4\pi \,\Delta t\,   a_{i+1/2}^{n}\,, \\
  r_{i+1/2}^{n+1}   = & \; r_{i+1/2}^{n} + 4\pi \,\Delta t\, v_{i+1/2}^{n} + 2\pi \,(\Delta t )^2 a_{i+1/2}^{n} \,,
\end{align}    
where $\Theta [\cdot] $ is the Heavyside function, $\Delta t$ is the time interval between the steps $n$ and $n+1$, $\Delta m_{i+1/2}$ is the mass at the interface, defined as a function of the mass of the two bounding shells $\Delta m_{i+1/2}=(\Delta m_{i+1}+\Delta m_{i})/2$, $\eta_{\rm vnr} = 2$ (the standard von Neumann-Richtmyer method) and $H$ is the viscous pressure \citep{Schulz:1964}.
The radius of each shell is defined as the barycenter radius:
\begin{align}  
r^n_i = \left(\frac{ (r_{i+1/2}^n)^3+  (r_{i-1/2}^n)^3}{2}\right)^{1/3}\,,
\end{align}  
and its density:
\begin{align}  
\rho^n_i = \frac{3 \Delta m_{i}}{4\pi(  (r_{i+1/2}^n)^3-  (r_{i-1/2}^n)^3)}\,.
\end{align}  
Instead, the pressure $p_i$ in the shell is derived by solving (either by successive iterations or by direct analytic solution in the special case of an ideal gas) 
%\dft{what do you mean by direct solution?}
%\rb{It means that, while a numerical iterative technique is required and used for a general EoS, in the special case of a ideal gas an analytic solution is found. The numerical implementation is described in Mezzacappa \& Bruenn 1993.}
the following equation for the specific internal energy $e_i$:
\begin{align}
e^{n+1}_i &= e^{n}_i -\frac{p^{n+1}_i  +p^{n}_i}{2} \left(\frac{1}{\rho^{n+1}_i }-\frac{1}{\rho^n_i } \right)\nonumber+\\
		& \quad-2\pi \Delta t \left(r^{n+1}_i  +r^{n}_i\right)^2 \,H_i^n \, \frac{v_{i+1/2}^{n}-v_{i-1/2}^{n}}{\Delta m_{i}}\,,
\end{align}
and assuming the following equation of state $p_i =2 \rho_i e_i/3$, appropriate for a perfect gas of adiabatic index $\Gamma = 5/3$. 
As stated before, the values of the quantities for the first interface are set, at the beginning of the lagrangian evolution, to those provided by the thin-shell model at $t_{\rm beg rev}$, by identifying  $r_{1/2} = R_{\rm pwn}$, $v_{1/2} = \dot{R}_{\rm pwn}$, $\Delta m_{1/2} = {M}_{\rm sh}$, and $p_{1/2} = P_{\rm pwn}$.

In practice, according to our treatment of the evolution during the reverberation phase, the PWN acts as a piston inside the SNR, in the same way as it was done for the lagrangian simulations in \citet{Bandiera:2023}, except that now the piston includes the swept-up shell, which provides its starting inertia and momentum. 
The mass of the swept-up shell is supposed not to change during the reverberation, which is equivalent to the assumption that no matter is further accumulated.
The evolution of $R_{\rm pwn}(t)$ is simply that of the inner interface of the first cell of the lagrangian algorithm.

The thin-shell model we used is the one of the TIDE code (see \cite{Martin_Torres:2022} and references therein), a one-zone time-dependent radiative code able to evolve the leptonic population of the PWN. 
TIDE solves the diffusion-loss equation considering adiabatic and radiative losses (synchrotron, IC, bremsstrahlung), plus diffusion (for a description of the equations see the following sections, and \cite{Martin_Torres+12a} for details of the radiative implementation).
We shall call the version of TIDE  as developed for this paper, TIDE+L, to easily distinguish it from thin-shell-only incarnation.
TIDE+L is based on the modification of TIDE already described in \citet{Bandiera:2020}, to which we add a number of modules to introduce the 
lagrangian treatment during reverberation, following the approach described before. 

\begin{comment}
We have developed a 1D Lagrangian hydrodynamical scheme, following the recipes described in \citet{Mezzacappa_Bruenn+93a}.
%
Shocks are handled through the implementation of standard von Neumann-Richtmyer viscosity \citep{Von-Neumann_Richtmyer50a}.

Here we introduce the new version of the TIDE code \citep{Martin_Torres+12a, Torres:2014, Martin_Torres:2022}, now coupled with our lagrangian code, described in our previous works in the series \citet{Bandiera:2021} and \citet{Bandiera:2023} (hereafter PaperI and Paper II, respectively). 
%
From now on we will call the present version of TIDE as TIDE+L, to easily distinguish it from thin-shell only versions.
%
TIDE+L allows to move through and beyond the reverberation phase, accounting for all energy losses the system might experience.
\end{comment}

%%%%%%%%%%%%%%%%%%%%%%%%%%%%%%%%%%%%%%%%%%%%%%%%%%
\section{Models}
\label{sec:selsources}
%%%%%%%%%%%%%%%%%%%%%%%%%%%%%%%%%%%%%%%%%%%%%%%%%%

In order to assess the performance of this new approach, we have selected a few representative cases for the evolution of the PWN-SNR system.
We shall compare the results for these cases with those obtained via a full-lagrangian evolution.

\begin{figure} 
\centering
\includegraphics[width=.48\textwidth]{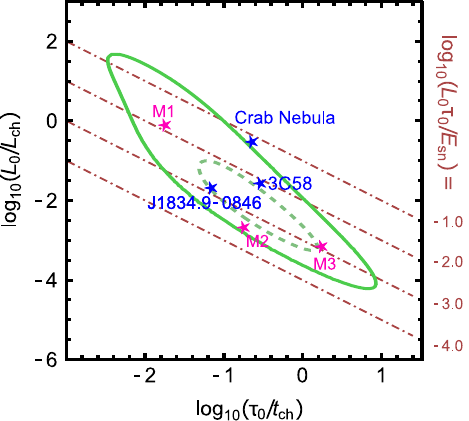}
        \caption{Plot of the PWNe population in the characteristic plane $\log_{10}{(\tau_0/\tch)}- \log_{10}{(L_0/\Lch)}$, as defined in \citet[][, Paper II]{Bandiera:2023}. The green solid line shows the iso-level enclosing the 98\% of the population, while the dashed one the 50\%.
        The position of the sources selected as references for the present study is shown with a star symbol: blue stars for real sources, magenta ones for synthetic sources.}
\label{fig:PAR+sources}	
\end{figure}

As it was pointed out in our previous works, all quantities can be scaled with the SNR characteristic units defined, following \citet{Truelove1999}, as: 
\begin{eqnarray}
\Rch\!\!\!&=&\!\!\!\Mej^{\,1/3}\rhoism^{\,-1/3},	\\
\tch\!\!\!&=&\!\!\!\Esn^{\,-1/2}\Mej^{\,5/6}\rhoism^{\,-1/3},
\label{eq:chscales}
\end{eqnarray}
where $\rhoism=1.4 \,m\rs{{p}} n\rs{ism}$ is the mass density of the ISM, with $m_{\mathrm{p}}$ the proton mass, and $n\rs{ism}$ the number density.
As shown in Paper II, making use of this scaling each PWN-SNR system can be represented by a point in the $\log_{10}{(\tau_0/\tch)}- \log_{10}{(L_0/\Lch)}$ characteristic plane{\color{black}, where $\Lch = \Esn/\tch$.}
The entire PWN-SNR population covers a roughly elliptic area in the same plane, as can be seen from Fig~\ref{fig:PAR+sources}.

Our benchmark models have been selected to  range over this region, in order to sample diverse properties of the reverberation phase.
They consist of three real sources plus three synthetic ones,  (see Fig.~\ref{fig:PAR+sources})
The three real sources are: the Crab nebula, 3C58 (both well known young systems, still in their free-expansion phase, with very detailed spectral data), and J1834.9--0846 (with a rather poorly covered spectrum but representative of a highly  compressible system). 
Since the scope of this work is to validate our new approach, and not to recompute 
spectral fittings, we adopt the same parameters as in Paper I and \citet{Martin_Torres:2022} for 3C58.
For synthetic sources we assume the same spectral parameters of the Crab nebula,  which are needed to compute the dynamical evolution of the system when radiative losses are included.
%
%%%%%%%%%%%%%%%%%%%%%%%%%%%%%%%%%%%%%%%%%
%%%%%%%%%%%%%%%%%%%%%%%%%%%%%%%%%%%%%%%%%
\begin{comment}
The population is represented (as a green solid line) in Fig.~\ref{fig:PAR+sources}, and the positioning of our benchmark sources is shown with colored stars (blue for real sources, magenta for synthetic ones).
\end{comment}
%
 %%%% Table with parameters for CRAB AND J1834.9 FROM BANDIERA 2020 (REVI)
 \begin{table*}
%\scriptsize
\centering
\caption{Parameters used in the simulations for the Crab Nebula, J1834.9--0846 and 3C58, as taken from Paper I and \citet{Martin_Torres:2022}. 
For J1834.9--0846, the unique source which reverberation phase has already started at its actual age, the original best-fit spectral parameters are no more able to fit the spectral data, then we show in parentheses the value of the magnetic fraction needed to get a good one. 
{\color{black}In the last row we also report the values of the magnetic field in the PWNe that we obtain at the actual age for the three real sources. }%Please notice that, as it is common with one-zone models, these values tend to underestimate those inferred by observations.}
\label{tab:pwnpar}}
\begin{tabular}{|lccccccc|}
\hline
Parameter & Symbol & Crab PWN & J1834.9--0846 & 3C58 & M1 & M2 & M3\\
 &   & \scriptsize{\textit{real}} & \scriptsize{\textit{real}} & \scriptsize{\textit{real}} & \scriptsize{\textit{synthetic}} & \scriptsize{\textit{synthetic}} & \scriptsize{\textit{synthetic}}\\
\hline
\rowcolor{lightgray} Observed Properties &  &  & & & & & \\
\hline
Spin-down period (s) & $P$ & 0.0334 & 2.48 & 0.0657 & -- & -- & --\\
Period derivative  & $\dot{P}$& $4.2\times10^{-13}$ & $7.96\times10^{-13}$ & $1.93\times10^{-13}$ &  -- & -- & --\\
Age (yr) & $t\rs{age}$ & 969 & 7970 & 2500 & -- & -- & --\\
Characteristic age (yr) & $\tau_c$ & 1296 & 4900 & 5398 & -- & -- & --\\
Distance (kpc) & $d$ & 2 & 4 & 2 & -- & -- & --\\
\hline
\rowcolor{lightgray}Pulsar Injection &  &  & & & & & \\
\hline
Braking index & $n$ & 2.51 & 2.2 & 3.0 & 2.33 & 2.33 & 2.33\\
Initial spin-down time (yr) & $\tau_0$ & 758 & 280 & 2878 & 220 & 1000 & 2500\\
Initial spin-down luminosity (erg s$^{-1}$) & $L_0$ & $3 
\times 10^{39}$ & $1.74 \times 10^{38}$ & $9.3 \times 10^{37}$ & $2.13 \times 10^{39}$ & $1.2\times 10^{37}$ & $1.6\times 10^{37}$\\
\hline
\rowcolor{lightgray}Environment &  &  & & & & & \\
\hline
SN explosion energy ($\times 10^{51}$ erg) & $E\rs{sn}$ & $1$ & $1$ & $1$ & $1$ & 1 & 1\\
ISM density (particles/cm$^3$) & $n\rs{ism}$ & 0.5 & 0.5 & 0.1 & 0.02 & 0.5 & 5.0\\
SNR ejected mass (M$_\odot$) & $\Mej$ & 9.0 & 11.3 & 17.2 & 12.0 & 16.9 & 8.0\\
\hline
\rowcolor{lightgray}Spectral model &  &  & & & & & \\
\hline
Far infrared temperature (K) & $T\rs{fir}$ & 70 & 25 & 25 & 70 & 70 & 70\\
Far infrared energy density (eV cm$^{-3}$) & $w\rs{fir}$ & 0.1 & 0.5 & 0.22 & 0.1 & 0.1 & 0.1 \\
Near infrared temperature (K) & $T\rs{nir}$ & 5000 & 3000 & 2900 & 5000 & 5000 & 5000 \\
Near infrared energy density (eV cm$^{-3}$) & $w\rs{nir}$ & 0.3 & 1 & 0.41 & 0.3 & 0.3 & 0.3\\
Energy break & $\gmbreak$ & $9 \times 10^{5}$ & $10^{7}$ & $8.8 \times 10^{4}$ & $5 \times 10^{5}$ & $5 \times 10^{5}$ & $5 \times 10^{5}$\\
Low energy index & $\alL$ & 1.5 & 1 & 1 & 1.5 & 1.5 & 1.5\\
High energy index & $\alH$ & 2.54 & 2.1 & 3.012 & 2.5 & 2.5 & 2.5\\
Containment factor & $\epsilon$ & 0.27 & 0.6 & 0.5 & 0.27 & 0.27 & 0.27\\
Magnetic fraction & $\eta$ & 0.02 & 0.045 (0.15) & 0.0106 & 0.02 & 0.02 & 0.02\\
\hline
\rowcolor{lightgray}Characteristic values &  &  & & & & &  \\
\hline
time (yr) & $\tch$ & 3329 & 4024 & 9766 & 12371.5 & 5623.4 & 1405.8\\
radius (pc) & $\Rch$ & 8 & 8.7 & 17 & 25.9 & 9.9 & 3.6\\
$\log_{10}{(\tau_0/\tch)}$ & $t^*$ & -0.64 & -1.16 & -0.53 & -1.75 & -0.75 & +0.25\\
$\log_{10}{(L_0\tch/\Esn)}$ & $L^*$ & -0.5 & -1.66 & -1.54 & -0.08 & -2.67 & -3.14\\
\hline
\rowcolor{lightgray}Derived quantities&  &  & & & & &  \\
\hline
Magnetic field at $\tage$ ($\mu$G) & $B\rs{PWN}$ & 100.0 & 2.2 & 2.0 & -- & -- & -- \\
\hline
\end{tabular}
\end{table*}
%%%%%%%%%%%%%%%%%%%%%%%%%%%%%%%%%%%%%%%%%%%%%%%
%
The relevant parameters defining the properties of the selected sources are given in Table~\ref{tab:pwnpar}. 
\begin{comment}
For the synthetic ones, the parameters are randomly generated once the positioning in the $\log_{10}{(\tau_0/\tch)}- \log_{10}{(L_0/\Lch)}$ plane has been defined.
\end{comment}

%%%%%%%%%%%%%%%%%%%%%%%%%%%%%%%%%%%%%%%%%
\section{Dynamical evolution}
\label{sec:dynevo}
%%%%%%%%%%%%%%%%%%%%%%%%%%%%%%%%%%%%%%%%%

In this section we compare the long term evolution (up to $\sim\,8\tch$) of our benchmark sources computed using different approaches and codes, namely:
\begin{enumerate}
    \item the full, albeit non-radiative, lagrangian code, in its original PWN+SNR version introduced in Paper II;
    \item the herein-developed TIDE+L, a fully-radiative, hybrid lagrangian code
     \item the new code TIDE+L, but used in the non-radiative regime simply obtained  by
     neglecting all radiative and diffusive losses. 
     To distinguish it from the radiative TIDE+L, we will refer to the no-radiative / no-losses version as to TIDE+L$^{\mathrm{NL}}$;
\end{enumerate}
The %ad-hoc 
introduction of TIDE+L$^{\mathrm{NL}}$ allows us to make a direct comparison with the results of the pure lagrangian approach, which is also non-radiative.
It serves to cross calibrate the dynamical part contained in TIDE, and gain confidence when the results are modified by the further
appearance of the radiation losses.
\begin{figure*} 
\centering
	\includegraphics[width=.48\textwidth]{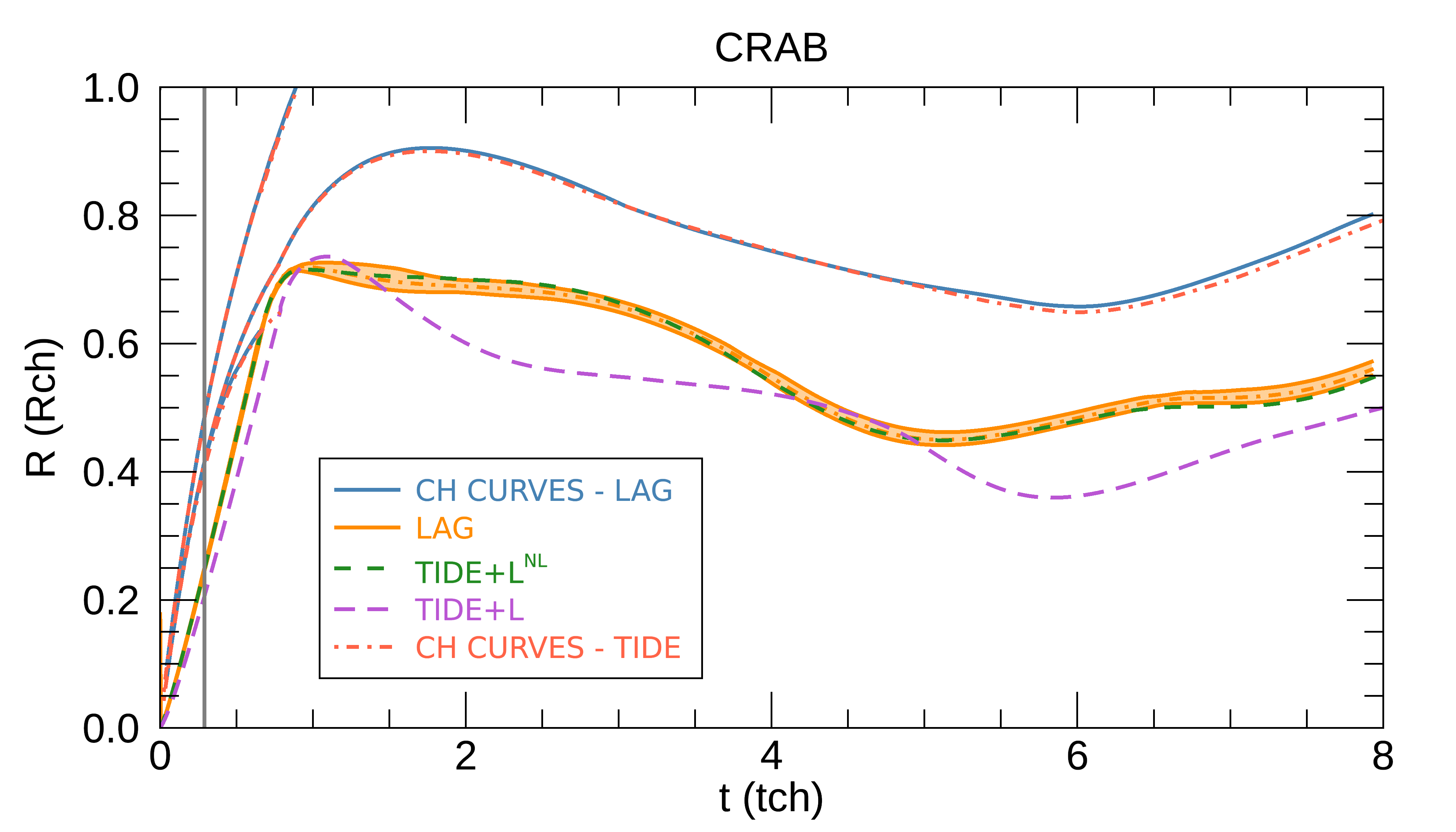}
    \includegraphics[width=.48\textwidth]{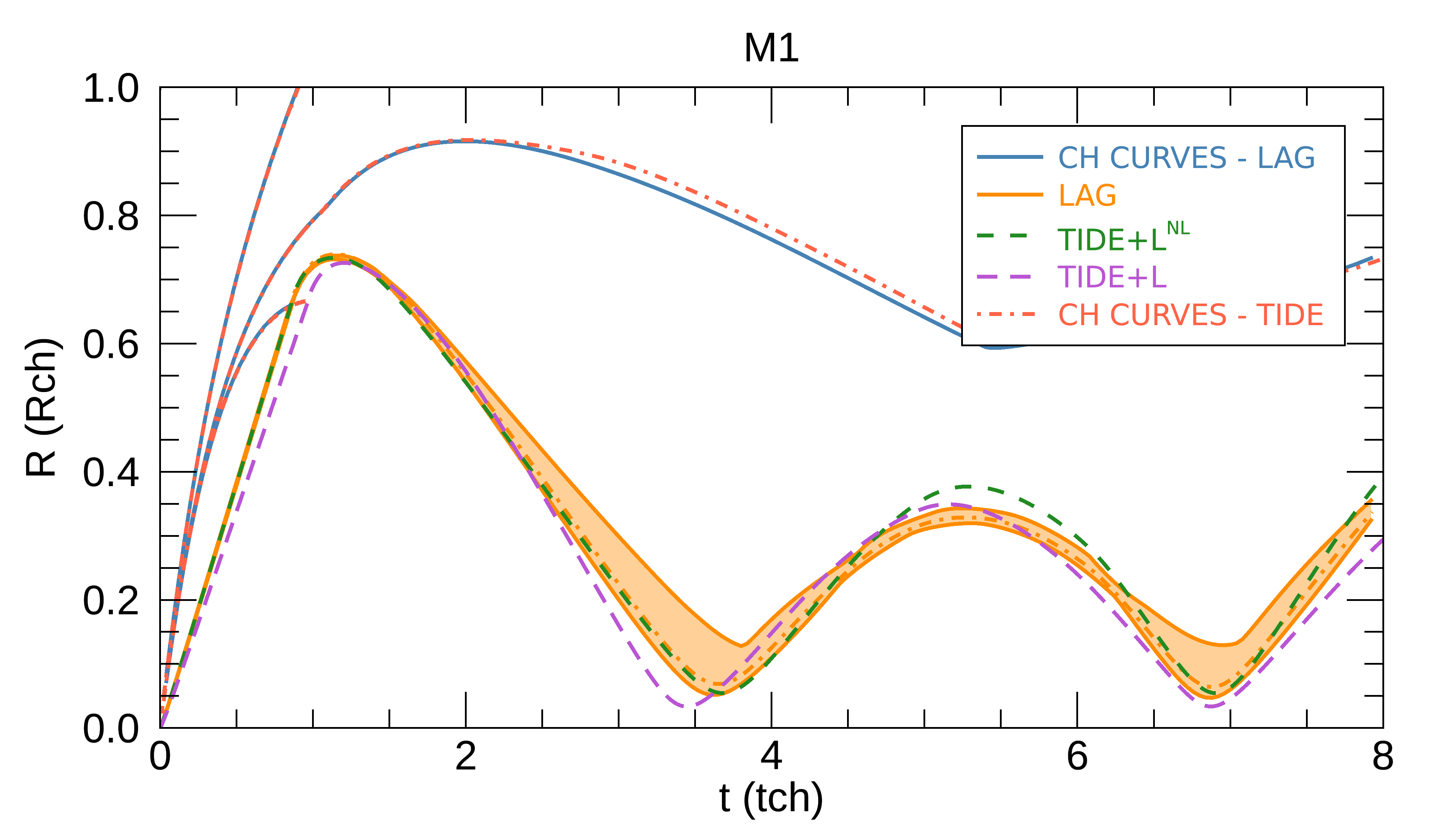}\\
     \includegraphics[width=.48\textwidth]{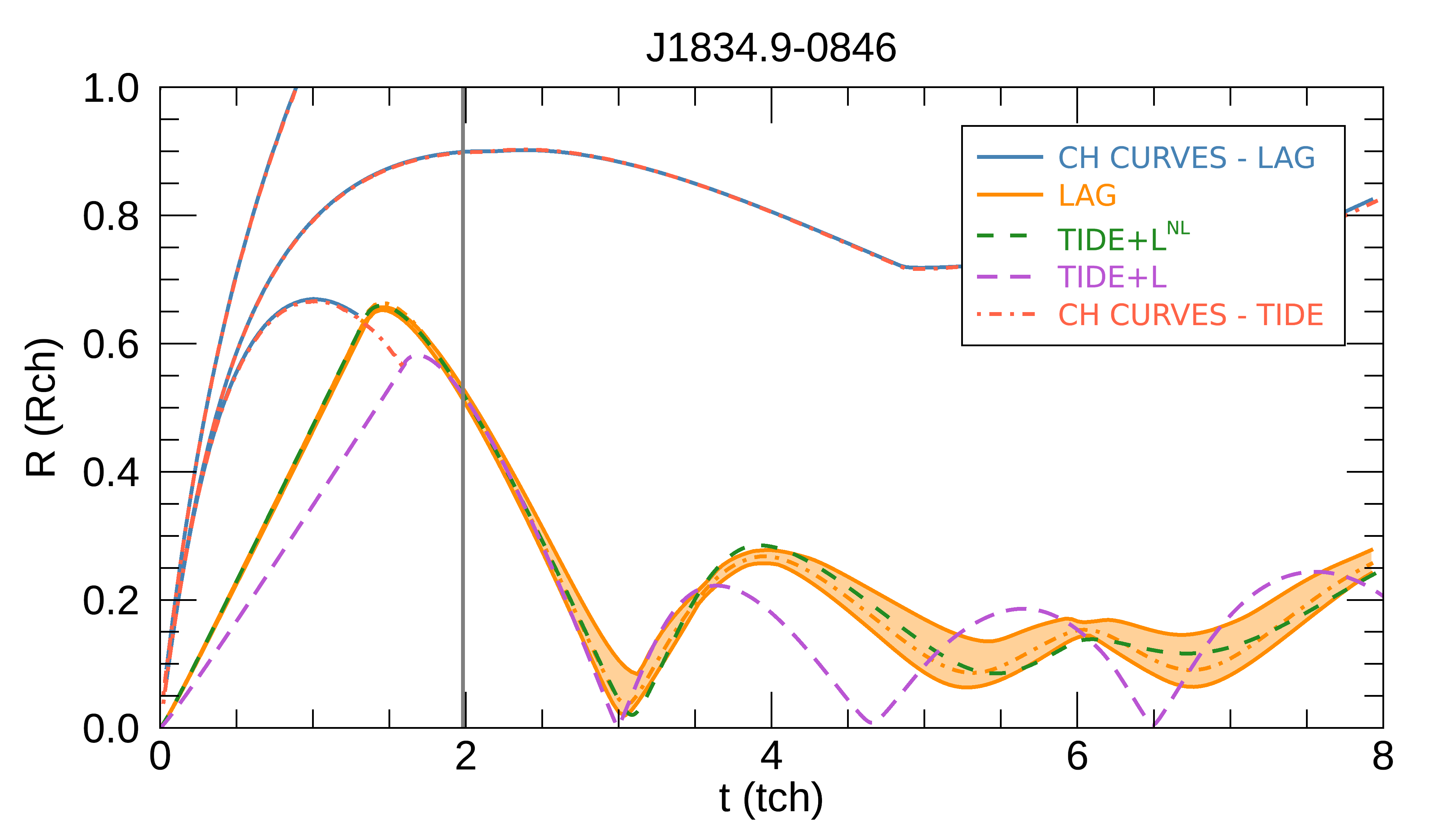}
     \includegraphics[width=.48\textwidth]{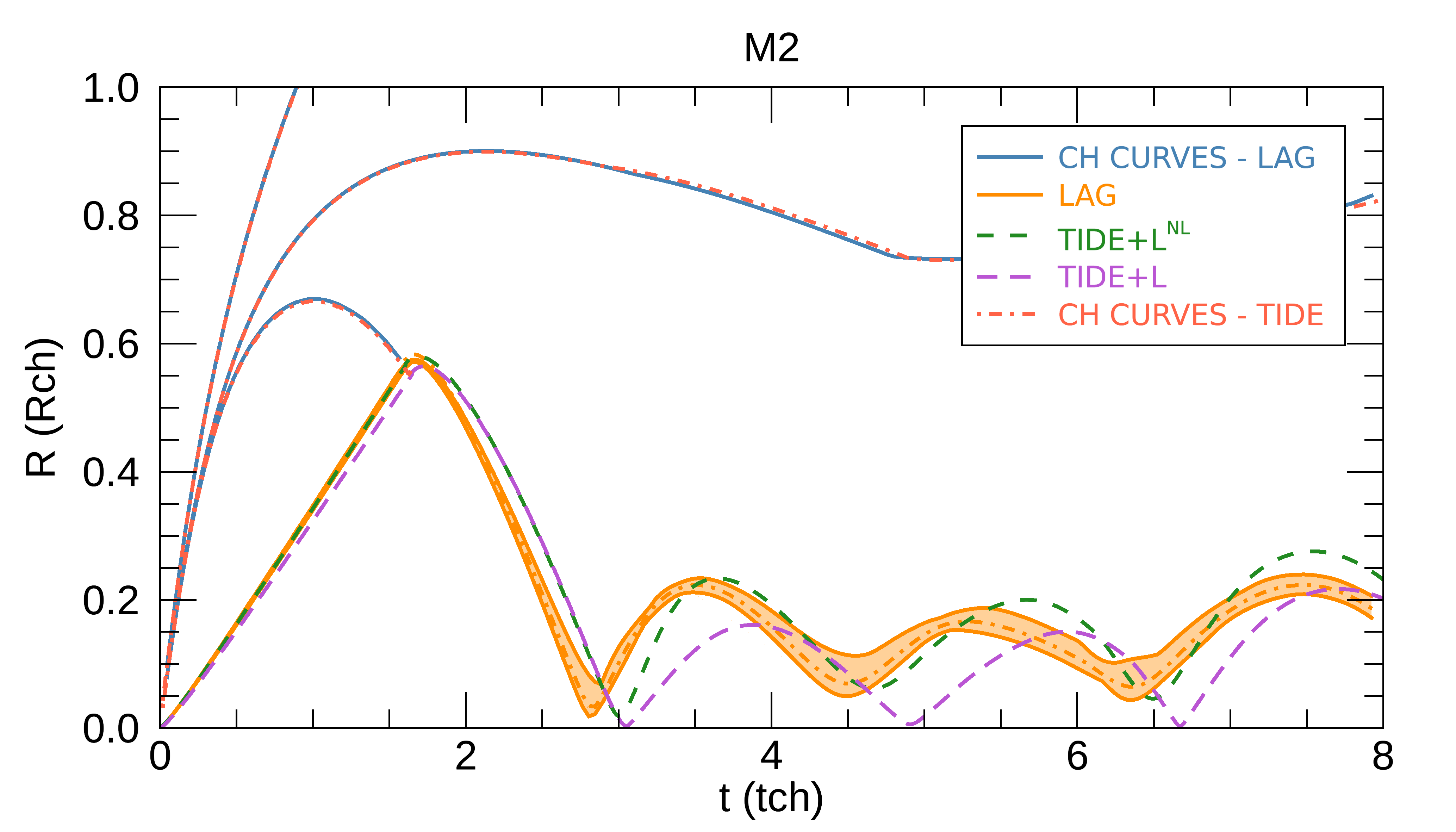}\\
     \includegraphics[width=.48\textwidth]{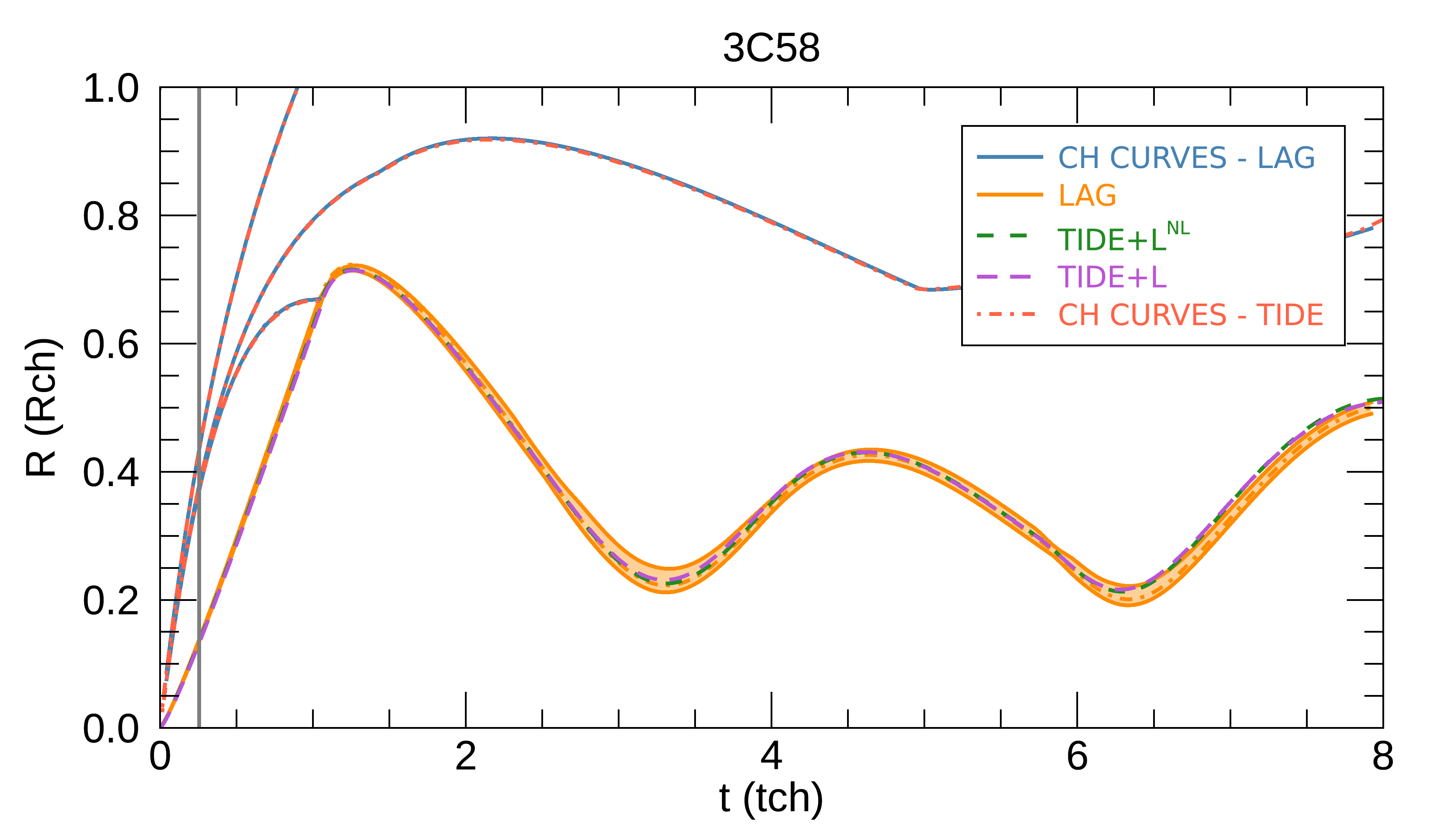}     
     \includegraphics[width=.48\textwidth]{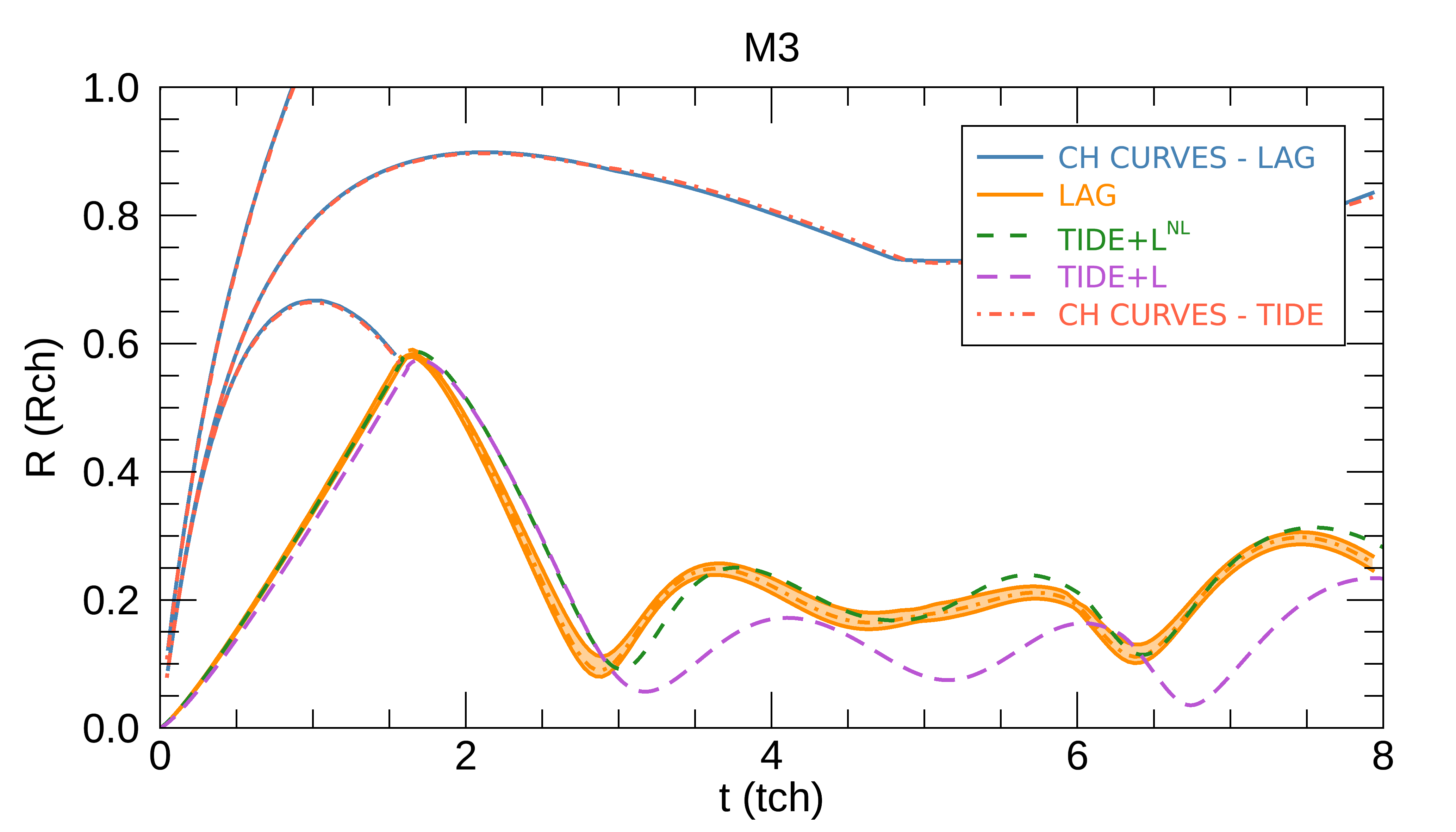}\\
        \caption{Time evolution of the PWN radius and SNR characteristic curves (named CH CURVES in the legend) for different models, with quantities expressed in characteristic units (see Eq.~\ref{eq:chscales}). The SNR FS, CD and RS are extracted from the lagrangian simulations (light blue solid lines) and compared with the same curves as obtained with TIDE+L, in the non radiative case (dot-dashed light red lines). From the outside (left upper corner of each plot) to the inside, we can recognize the FS going out of the plot at $\sim1\tch$, the CD and the RS, impinging on the PWN. It is apparent that, except for minor deviations in the CD, the characteristic curves are perfectly coincident from the lagrangian code and TIDE+L. 
        From the lagrangian simulations we do not only plot $\Rpwn$, but also the outer edge of the swept-up mass shell (all in orange), to illustrate how much the shell deviates from the thin-shell approximation. The shell barycenter is also shown as an orange dot-dashed line lying in between the PWN radius and the outer edge of the shell.
        Green and violet dashed lines show the evolution of $\Rpwn$ as obtained, respectively, with TIDE+L$^{\mathrm{NL}}$ and TIDE+L. The presence of early synchrotron losses results in a deviation of the PWN radius already in the free-expansion phase and in the delayed beginning of the reverberation phase (particularly evident in J1834.9-0846).
        Each source is modelled following the parameters given in Table~\ref{tab:pwnpar}. Real sources are shown on the left column, synthetic ones on the column on the right.
        Finally, for real sources, we show the positioning of their present age on the evolutionary plot by means of a vertical solid gray line.}
\label{fig:EVOall}	
\end{figure*}
%%%%%%%%%%%%%%%%%%%%%%%%%%%%%%%%%%%%%%%%%

The results for the dynamical evolution of all benchmark sources are shown in
Fig.~\ref{fig:EVOall}. 
For each source, we present the time evolution of the PWN radius and of the SNR characteristic surfaces (the contact discontinuity, CD, the reverse shock, RS, and the forward shock, FS), all in terms of  characteristic units. 
%Solid lines always refer to the lagrangian code (LAG), dashed ones to TIDE+L. 
%
In order to assess the validity of the thin-shell approximation, we also show the swept-up shell thickness, as computed by the full-lagrangian code, where it is fully resolved.
\begin{comment}(as a filled orange colored area)
to appreciate the variation from the thin-shell approximation, that is especially apparent in some cases, as we will comment later.
As in our previous works of the series, the effect of the reverberation phase is measured through the compression factor of the PWN, namely:
\end{comment}

The strength of the reverberation phase is measured through the compression factor of the PWN, namely:
\begin{equation}
    \mathrm{CF}=\frac{\Rpwn^{\mathrm{max}}}{\Rpwn^{\mathrm{min}}}\,,
    \label{eq:CF}
\end{equation}
always computed at the first compression event. 
The PWN reaches its maximum radius $\Rpwn^{\mathrm{max}}$  close to the time of the beginning of the reverberation, $\tbegrev$, while $\Rpwn^{\mathrm{min}}$ is the PWN radius at the minimum of the first compression.
All these quantities are given in  Table~\ref{tab:CFs}, for the various approaches.
%
%%%% Table with CFs
\begin{table}
\footnotesize
\centering
\caption{CFs and related quantities as obtained, for the different sources considered in this work, with the pure lagrangian approach and the two versions of TIDE+L (the non radiative one indicated with TIDE+L$^{\mathrm{NL}}$). \label{tab:CFs}}
\begin{tabular}{llcccc}
\hline
  & &  & & &   \\
{\bf System} &  Code & $\tbegrev$ &  $\Rpwn^{\mathrm{MIN}}$ & $\Rpwn^{\mathrm{MAX}}$ & CF \\
&   & \scriptsize{[$\tch$]} & \scriptsize{[$\Rch$]} & \scriptsize{[$\Rch$]} & \\
\hline
\hline
%%%%% CRAB ---> ok
\multirow{4}{4em}{CRAB} & LAG  & 0.6852 & 0.4414 & 0.6385 &   1.447 \\ 
  & LAG(bary) & $\cdot$ & 0.4499 & 0.7205 & 1.602
  \\ 
  & TIDE+L$^{\mathrm{NL}}$  & 0.6408 & 0.4489 & 0.7153 & 1.593 \\ 
  & TIDE+L & 0.6633  & 0.3595 & 0.7357  & 2.047 \\ 
 \hline
 %%%%% J1834 ---> ok
 \multirow{4}{6em}{J1834.9--0846} & LAG  & 1.323 & 0.0198 & 0.6196 & 31.29 \\
  & LAG(bary) & $\cdot$ & 0.0368 & 0.6629& 18.01  \\ 
  & TIDE+L$^{\mathrm{NL}}$  & 1.315 & 0.0205 & 0.6592 & 32.16  \\
  & TIDE+L    & 1.339 & 0.0024  & 0.5819 &  242.5\\
 \hline
 %%%%% 3C58 ---> ok
 \multirow{4}{4em}{3C58} & LAG & 1.082 & 0.1915 & 0.6837  & 3.570 \\
   & LAG(bary)  & $\cdot$ & 0.2009 & 0.7232 & 3.600\\
  & TIDE+L$^{\mathrm{NL}}$  & 1.054 & 0.2128 & 0.7154  &  3.362   \\
  &  TIDE+L & 1.062  & 0.2161  & 0.7143 &  3.305\\
  \hline
  %%% M1 ---> ok
 \multirow{4}{4em}{M1}  & LAG  & 0.8806 & 0.0471 & 0.6746  & 14.32 \\
   & LAG(bary) &$\cdot$ & 0.0638 & 0.7394  & 11.59\\
  & TIDE+L$^{\mathrm{NL}}$ & 0.8466  &  0.0544  & 0.7339   & 13.49 \\ 
   & TIDE+L  & 0.8574 & 0.0335 & 0.7260 & 21.67  \\
 \hline
 %%%% M2 ---> ok
 \multirow{4}{4em}{M2} & LAG  &  1.604 & 0.0181  &   0.5618 & 31.04 \\
  & LAG(bary) & $\cdot$ & 0.0332 & 0.5829 &  17.56 \\
  & TIDE+L$^{\mathrm{NL}}$ & 1.592  &  0.0182   &  0.5799 & 31.86 \\
  & TIDE+L  &  1.606 & 0.0023 & 0.5645 & 245.4 \\
  %%%% M3  ----> ok
 \hline
 \multirow{4}{4em}{M3}  & LAG   & 1.5317  & 0.0801 & 0.5483 & 6.845 \\
  &  LAG(bary) & $\cdot$ & 0.0900 & 0.5905 & 6.561 \\  
  & TIDE+L$^{\mathrm{NL}}$  & 1.5393 & 0.0918 & 0.5877 & 6.402 \\
  &  TIDE+L &  1.5838 & 0.0355 & 0.5745  &  16.18\\  \hline
\end{tabular}
\end{table}
%%%%%%

%It is apparent that 
The difference in the compression factor is very small in all cases when we compare the evolution computed with the lagrangian code and  TIDE+L$^{\mathrm{NL}}$.
We found that for those systems who preserve a rather thin-shell after the onset of the reverberation phase (namely Crab, 3C58 and M3), the CF computed with TIDE+L$^{\mathrm{NL}}$ is also very similar to the one obtained considering the radius of the barycenter of the shell, computed as:
\begin{equation}
    \Rbary(t) = \frac{4\pi}{\Mswept}\int^{\Rshell(t)}_{\Rpwn(t)} \rho(r,t)\, r^3 dr\,, 
    \label{eq:bary}
\end{equation}
where $\Mswept$ is the swept-up mass collected in the shell at $t=\tbegrev$ and $\rho(r,t)$ the mass density.
This happens because in those cases the thin-shell approximation is particularly good, given the limited thickness of the shell; in fact we see that at the time the PWN reaches its minimum radius at first compression $\Rpwn^{\mathrm{MIN}}$, that we named $t\rs{MIN}$, $\Delta \Rshell(t\rs{MIN})/\Rpwn^{\mathrm{MIN}}=0.04,\,0.14,\,0.33$, for Crab, 3C58 and M3 respectively, is always less than unity.
%in these cases we have that at the time the PWN reaches its minimum radius at first compression $\Rpwn^{\mathrm{MIN}}$, that we named $t\rs{MIN}$, $\Delta \Rshell(t\rs{MIN})/\Rpwn^{\mathrm{MIN}}=0.04,\,0.14,\,0.33$, {\bf for Crab, 3C58 and M3, respectively, always less than unity}.
%
On the contrary this is no longer true when we look at the more  compressible  systems (namely J1834.9--0846, M1 and M2), for which the shell shows an  evidently larger extension with a relative thickness with respect to the minimum radius at first compression always larger than unity: $\Delta \Rshell(t\rs{MIN})/\Rpwn^{\mathrm{MIN}}=2.9,\,1.5,\,2.8$. 
Then the thin shell approximation is not valid.
Thus the value of the CF computed with the shell barycenter radius is no longer compatible with that obtained at the PWN boundary.

The SNR characteristic curves appear almost perfectly coincident between the lagrangian and TIDE+L results, with at most small variations in the CD when there is a larger deviation of the PWN radius after the first re-expansion caused by the presence of reflected shocks in between the PWN and the CD.

The changes introduced by the radiative losses  are 
evident for all systems, except for 3C58 that is characterized by a very low magnetization according to the fitting.
Losses reduce the energetics of the PWN since the very early free-expansion phase, leading to a slower expansion rate, and a delay of $\tbegrev$.
This is particularly evident for those systems that, like in the case of J1834.9--0846 or M1, have a short spin-down time $\tauz$ ($\ll \tch$) and a large initial magnetic fraction $\eta$.
%
%This is particularly evident, in the case of J1834.9--0846, due to the combination of its spin-down time $\tauz$ much smaller than the characteristic time %$\tch$, and the large initial magnetization. A similar case, even if its smaller magnetization leads to lower losses in the initial phase, is that of M1, %also having a $\tauz\ll \tch$ .
%
As expected, all cases (except 3C58) show a larger compression factor in the radiative regime and, 
\begin{comment}
depending on how much different the first compression is, this also 
\end{comment}
 this changes the subsequent 
evolution.

We thus notice that,
except for very compressible systems (e.g., J1834.9--0846 and M2), 
the variation of the CF induced by the radiative losses is rather modest, although nevertheless non-negliglible in general. 
The ratio between the CF computed from TIDE+L and TIDE+L$^{\mathrm{NL}}$ from Table 2 amounts to
a factor 1.28 for the Crab Nebula, 1.61 for M1 and 2.5 for M3, the less energetic of the three.
%a factor 1.41 for the Crab Nebula, 1.51 for M1, and 2.36 for M3, the less energetic of the three.
%\dft{changed numbers next, according to ratio between TIDE+L and Lag from Table 2, I could not reproduce which ratio you were quoting here before -- prior numbers are commented in the tex file} \BO{-- My fault, this is the ratio between TIDE+L and TIDE+L non radiative, to show the variation introduced by radiation losses.}
%
Essentially no variation is indeed observed for 3C58.
On the other hand,  more compressible systems show larger variations as it might be expected, with a factor of 7.5 for J1834.9--0846 and 7.7 for M2, at the boundary of the population.

One might wonder why there is such a big difference in the compression experienced by J1834.9--0846, M2 and M3, all characterized by a rather similar energetics ($L_0\tau_0/\Esn$) and close to the lower boundary of the population. 
This can be actually understood comparing the initial spin-down time with the characteristic age of the systems: M2 and J1834.9--0846 have $\tau_0/\tch\simeq 0.18$ and $\tau_0/\tch\simeq 0.07$ respectively, while M3 has $\tau_0/\tch\simeq 1.8$. 
This translates in a huge difference in the power that is still injected in the PWN at $\tbegrev$, namely: $3\times 10^{-4} \, L_0$ for J1834.9--0846, $3\times 10^{-3} \, L_0$ for M2 and $0.22 \, L_0$ for M3.
This last system then is still enough powerful to react to the SNR compression when reverberation starts, while the other two can barely contrast the pressure exerted by the SNR. This is reflected in the higher CF we found for J1834.9--0846 and M2.

Systems with large compression CF$\,\gg 100$ are possible but rare, limited to the extremes of the known population of PWNe. In these, there will be extreme modifications of the PWN spectrum as time goes by, with significant increase of magnetic field and burn-out of electrons, as well as the appearance of super-efficiency, {\color{black} a phase characterized by an  emitted luminosity at a given energy band that exceeds the pulsar spin-down power at the same age} \citep{Torres_Lin:2018}.

In our non-radiative study of Paper II we have produced maps of $\log_{10}(\mathrm{CF})$ in the then-considered parameter space, respectively using a thin-shell model with different prescriptions for {\color{black} the pressure outside the PWN} or our interpolating formula to our lagrangian models. 
We refrain to do so here since the problem is complicated by the existence of radiation, that is, the values of $\log_{10}(\mathrm{CF})$ not only should account for the 
dynamical and energetics features, but also for the assumed magnetization. It is not the aim of this work to do a full population analysis or a phase-space exploration of all combinations we could think are plausible in nature.

{\color{black} Radiative losses induce an extra-compression that is very mild for most of the population. On the other hand there is limited sector of the population, namely the very low energetic systems, in which losses can lead to a substantial enhancement (up to 10 times) of the compression.}
%The extra-compression induced by the radiative losses during reverberation is very mild for most of the population,
%while it might enhance the CF by a maximum factor of $\sim 10$ only for a limited sector of the population, that of very low energetic systems. 
%

Finally, as for comparison with the present results, we want to recall the CFs we estimated in Paper I for the Crab nebula and J1834.9--0846, using the standard version of the TIDE code without the lagrangian module
%(version 2.3) 
that is based on the pure thin-shell approximation (similarly to all the other one-zone models available so far in the literature). 
What we obtained then was: a CF of 3.530 for the Crab (2.712 with no radiative losses); a CF of 1054 (80.75 with no losses) for J1834.9--0846.
The difference with the present results, which maps very well what is expected from the lagrangian evolution, is apparent: the pure thin-shell approximation always leads to an overestimation of the CF (by a factor 1.7 for the Crab, 4.3 for J1834.9--0846), as we extensively discussed in Paper II.
As a general result, we have shown here that the CF is very sensitive to the details of the model used, so that an accurate treatment is a must, to avoid inaccurate results. 
%\dft{commented 'grossly' since in fact, to me, erring by a factor of 2 to 5 in the extreme of the population is actually quite mild, for them, yes,  we do not have CFs of 1000 but we do have compressions of 250 even in the full lagrangian scheme, which is indeed quite impressive a fact (is there any other extended doing this change of bouncing dynamical scale? Stars, yes, but they are not extended...) [This is probably something to say if we indeed think there is no other system doing it]. I suggest to add this for the moment:} 
Despite the corrections introduced by the better dynamical treatment, which by themselves represent {\color{black}an important} progress in PWNe simulations, providing the tools to pass through reverberation in a relatively safe approach, the compression factors at the border of the known population of nebulae remain extreme. PWNe can reduce themselves in size by more than two orders of magnitude, at least in the 1D representation, while in 3D this behaviour might change due to the complex interaction of the PWN boundary with the SNR and the onset of border instabilities and mixing. The main spectral evolution effects of such changes are discussed next, with further details left for {\color{black} future} discussion elsewhere.

%%%%%%%%%%%%%%%%%%%%%%%%%%%%%%%%%%%%%%%%%%%%%%%%%%
\section{Spectral evolution}
\label{sec:tide-nonrad-rad}
%%%%%%%%%%%%%%%%%%%%%%%%%%%%%%%%%%%%%%%%%%%%%%%%%%%%

Let us discuss here the spectral  evolution of our three real sources, as computed by TIDE+L.
For all sources we compute the total spectrum at their present age, to be compared with available spectral data. 
We then perform a multi-epoch analysis to investigate how, especially in %salient 
relevant moments during the reverberation phase, the spectral properties of the sources may vary.

As we already mentioned before, realizing a good fit to the data, i.e. one that takes into account the modification TIDE+L brings over previous versions regarding the lagrangian implementation of the evolution, is beyond the scope of the present work. We leave this to future research.
Rather, we here give preference to the comparison with the results obtained with the previous -- thin-shell -- version of TIDE. 
It is in this comparison where we shall be able to
learn the impact of incorporating the new treatment of the dynamics via the lagrangian modules that represent the passage through reverberation.
{\color{black} For this reason we have not modified the spectral parameters determined in the reference works for the selected sources, not to introduce differences in the spectral properties besides those naturally arising from the different evolution. For a discussion of the fitting procedure and best-fit values we refer to Paper I (and references therein) and  \citet{Martin_Torres:2022}.}

\begin{comment}
    We have then not modified the spectral parameters from those reported in Table~\ref{tab:pwnpar}. 
\end{comment}
%
For the Crab and 3C58, even without optimizing the parameters,
the fits appear to be still good, given that both   are still in free-expansion, and the thin-shell approximation provides a good description of their evolution, such that the difference between TIDE+L and previous versions, are small. 
Different is the case of  J1834.9--0846, which  is the only source that is already in reverberation, according to our models (see \citealt{Torres:2017}).
%%%%%%%%%%%%%%%%%%%%%%%%%%%%%%%%%%%%%%%%%
\begin{figure*} 
\centering
	\includegraphics[width=.49\textwidth]{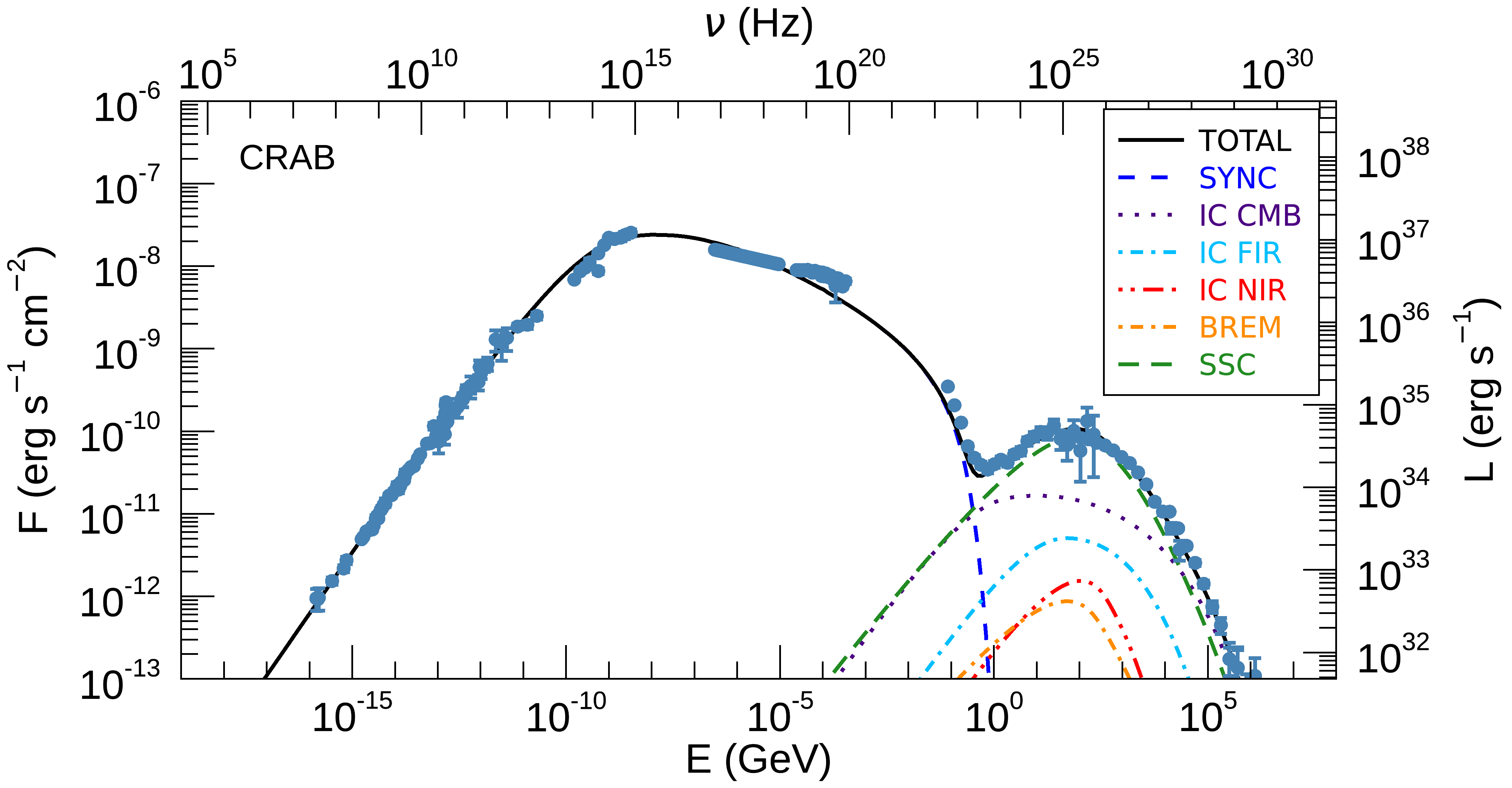}
 \hspace{0cm}
 	\includegraphics[width=.49\textwidth]{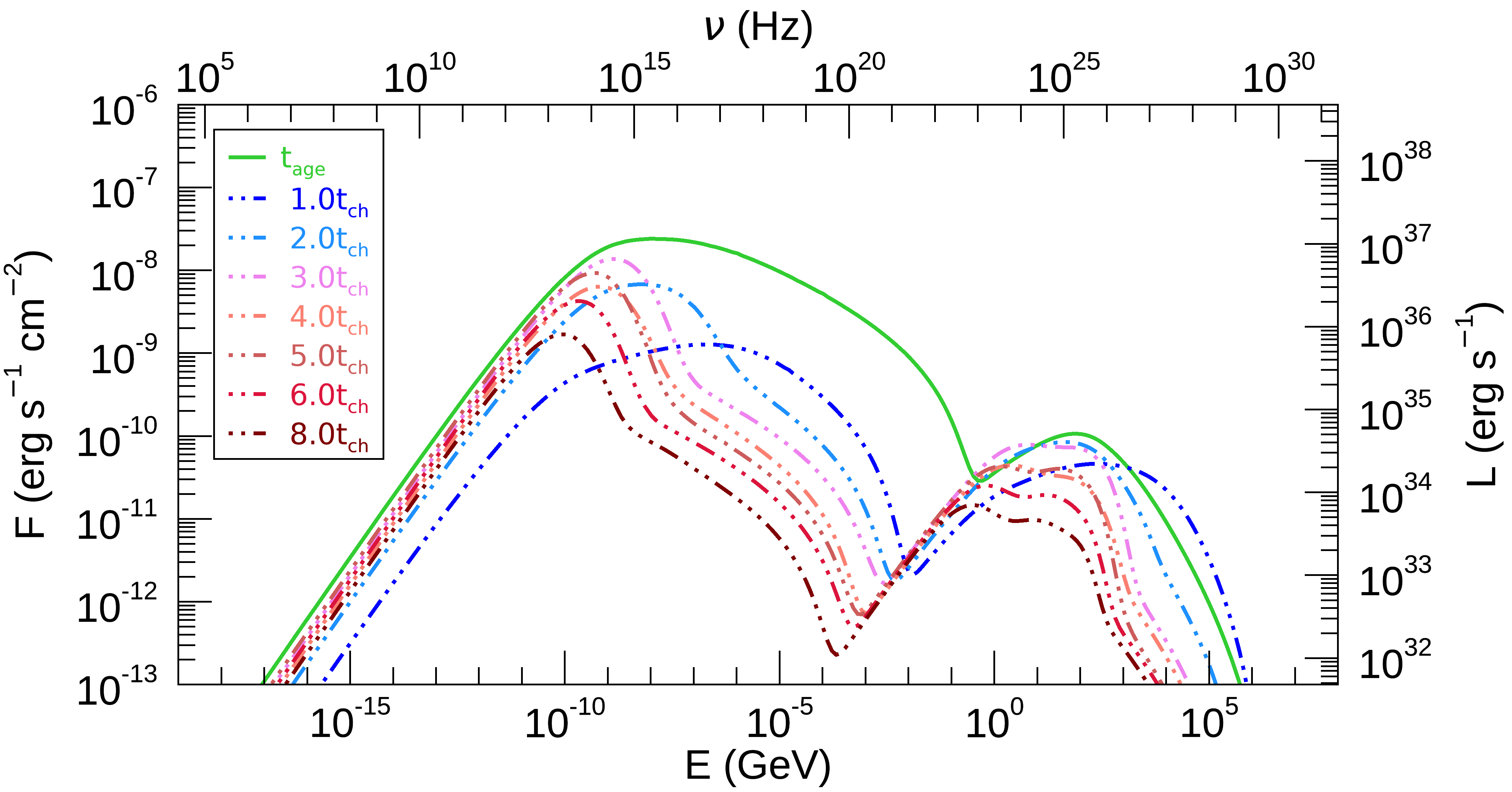}\\
\vspace{0.1cm}
\includegraphics[width=.49\textwidth]{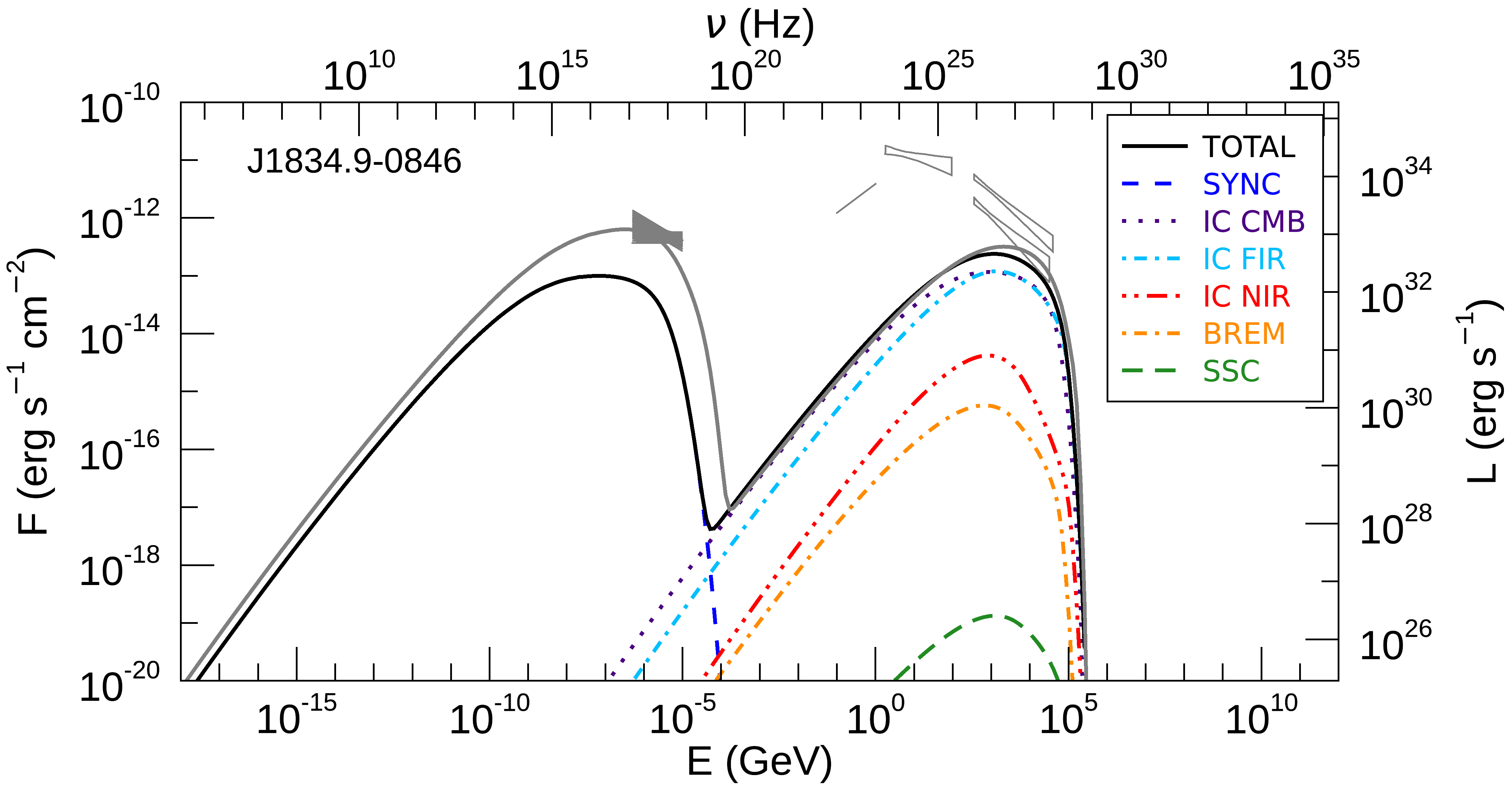}
 \hspace{0cm}
 	\includegraphics[width=.49\textwidth]{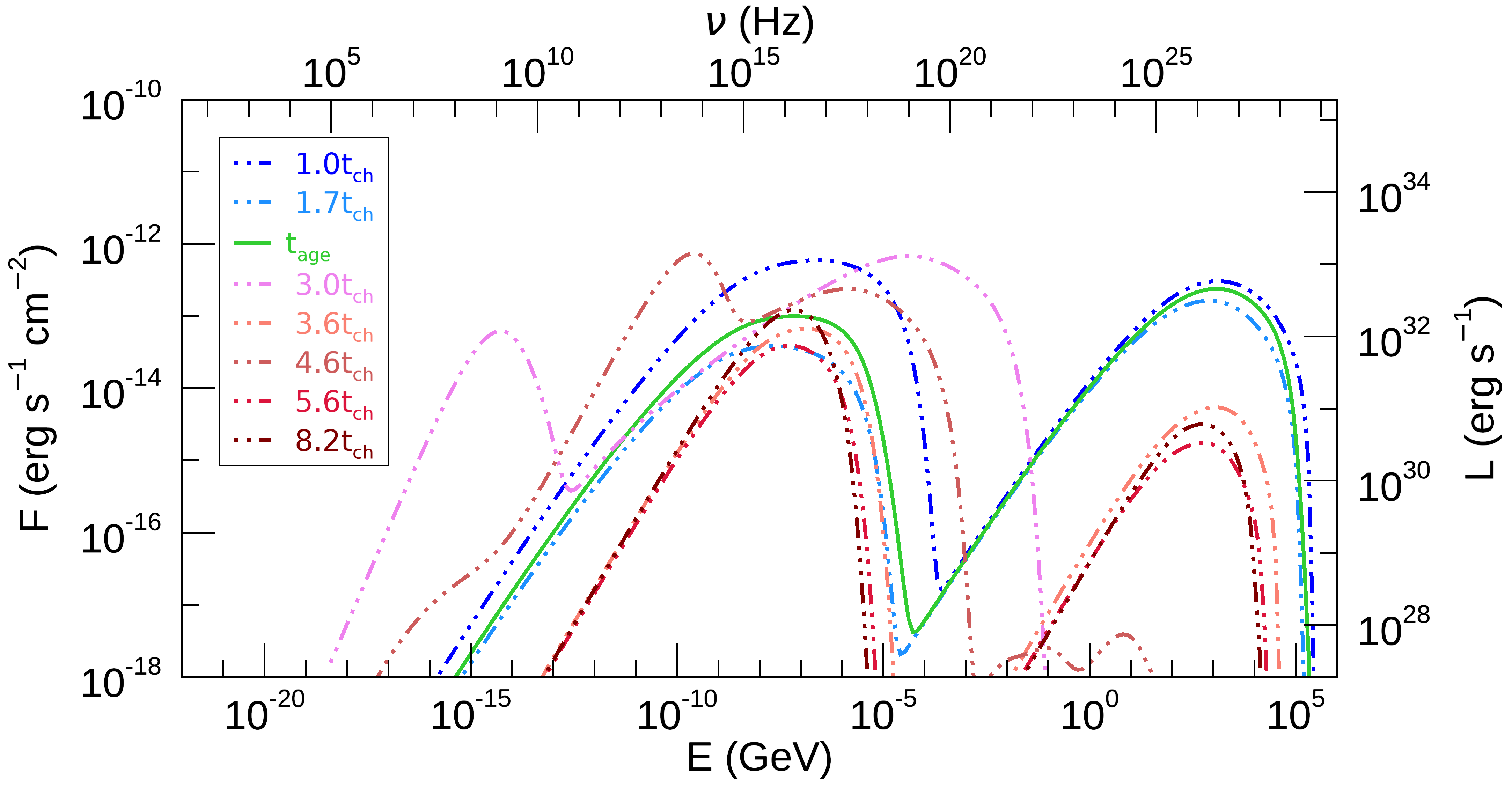}\\
\vspace{0.1cm}
  	\includegraphics[width=.49\textwidth]{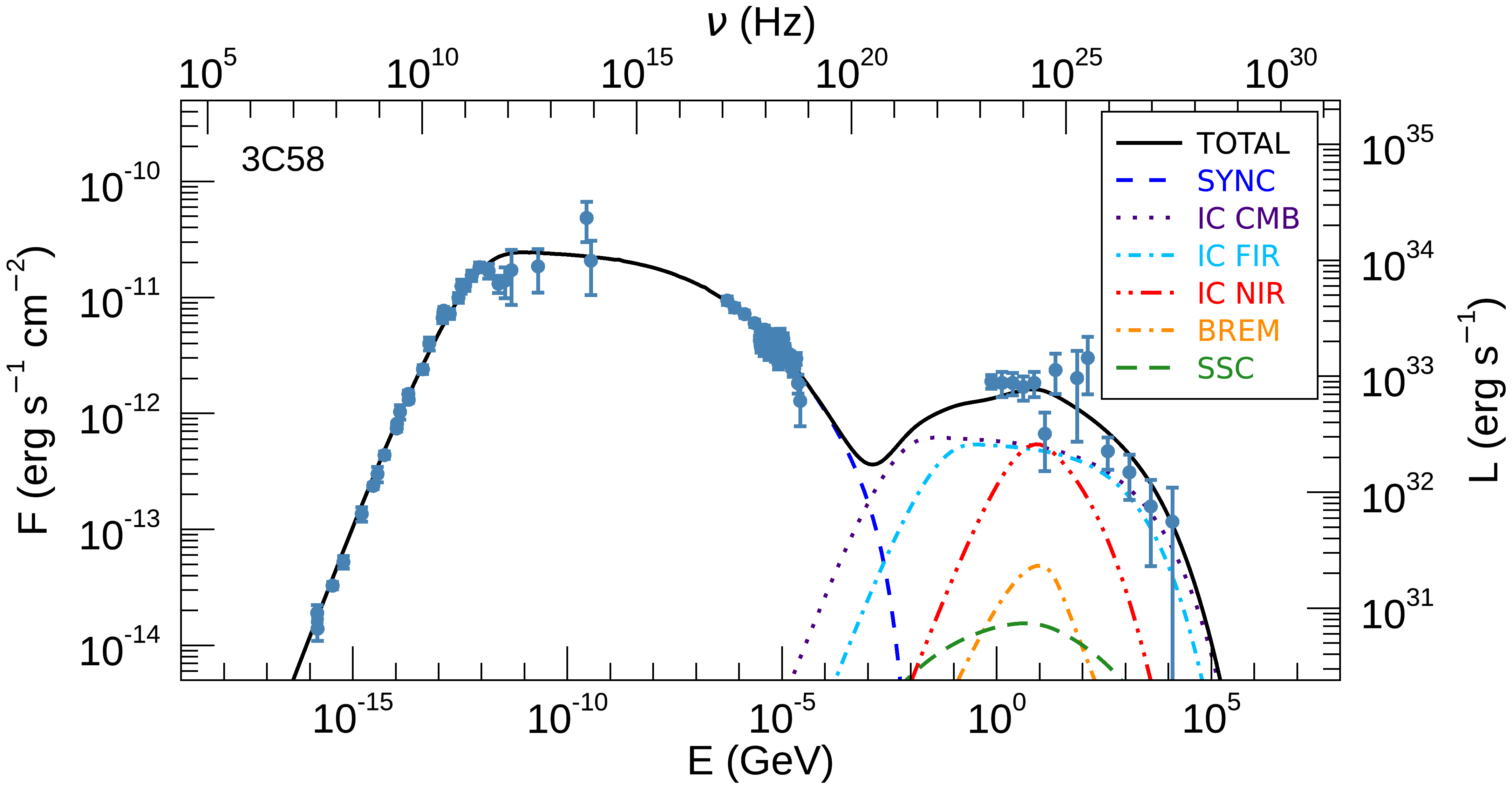}
 \hspace{0cm}
 	\includegraphics[width=.49\textwidth]{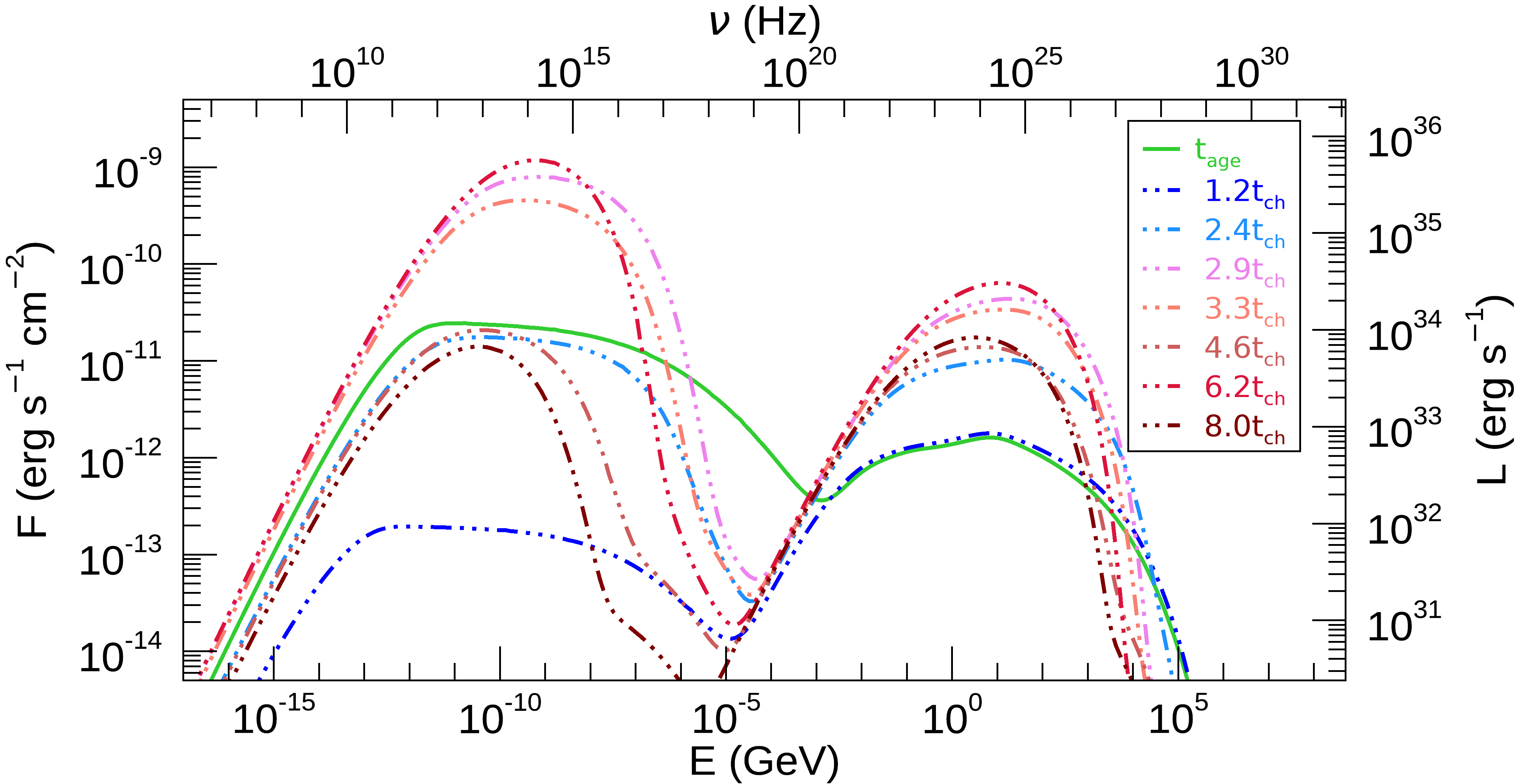}
  \caption{Spectra at $\tage$ (panels on the left) and at multiple epochs (panels on the right) for the three real benchmark sources considered in this work (from top to bottom: Crab, J1834.9--0846 and 3C58). 
  The single spectral components contributing to the total emission are plotted separately at $\tage$, where SYNC stands for synchrotron emission from both particles families, SSC for self-synchrotron Compton and the major photon fields for IC considered are: the cosmic microwave background (CMB), far and near infrared photons (FIR and NIR).
  In the multi-epoch spectra the spectrum at $\tage$ is plotted again for an easier comparison with other ages, in green color and solid line. The different epochs considered, chosen in order to map salient moments of the evolution of each source (e.g. maximum or minimum of $\Rpwn$), are listed in the plot legend, and relative spectra are drawn with the same color-code.
  For the Crab Nebula data are taken from \citet{Baldwin:1971, Macias-Perez:2010} (radio), \citet{Buhler:2014} (from IR to very high energies) and \citet{LHAASO_Crab:2021} for extremely high energy data.
  For J1834.9--0846 we consider the same set of data as in \citet{Torres:2017}: out of them only X-ray data are certainly referring to J1834.9--0846 and coming from the magnetar nebula \citep{Younes:2016}, while in the GeV and TeV bands we can only report upper limits. As mentioned in the main text, since this source at $t=\tage$ already entered its reverberation phase, its spectral properties are particularly sensitive on the variation of the PWN radius, pressure and magnetic field. This is the reason why in order to get a reasonable fit of the data we need to modify substantially the initial magnetization from the original value assumed in the reference work using the pure thin-shell TIDE ($\eta=0.045$ vs the new value $\eta=0.15$, much larger). The total spectrum corresponding to the new best-fit value is shown as a gray solid line.
  Data for 3C58 come from \citet{Green:1986, Morsi:1987,Salter:1989, Torii:2000} (radio and IR) \citep{Green:1994,Slane:2008} (X-rays), \citet{Abdo:2013,Ackermann:2013,Li:2018} (gamma-rays) and \citet{Aleksic:2014} (for the very high energies). } 
\label{fig:all_sp}	
\end{figure*}

Here we briefly recall how we compute the spectral model through the source evolution. 
The energy input, {\color{black} namely }the spin-down luminosity of the pulsar $L(t)$, changes with time following:
\begin{equation}\label{eq:spin-down}
    L(t) = L_0 \left( 1 + \frac{t}{\tau_0} \right)^{-\beta}
\end{equation}
where $\beta=(n+1)/(n-1)$ is the fading index.
The particle content of the PWN  at each time is given by the balance of injection, energy losses and adiabatic losses (or gains, in compression phases), plus possibly the escape of particles from the source through diffusion.
These are taken into account solving the following diffusion-loss equation, as used in \cite{Martin_Torres+12a,Torres:2014}:
\begin{equation}\label{eq:diffloss}
    \frac{\partial N(\gm,t)}{\partial t} = -\frac{\partial}{\partial {\gamma}}\left[ \dot{\gm}(\gm,t) N(\gm,t)\right] - \frac{N(\gm,t)}{\tau(\gm,t)} + \Qinj(\gm,t)\,,
\end{equation}
where $N(\gm,t)$ is the particle number, $\gm=E/(m_e c^2)$ the particle normalized energy (with $m_e$ the electron mass, considering a fully leptonic pulsar wind), $\dot{\gm}(\gm,t)$ the loss (gain) term (containing synchrotron, IC and bremsstrahlung radiative losses plus adiabatic losses -- or gains), $\tau(\gm,t)$ is the escape term (assuming Bohm diffusion) and, finally, $\Qinj(\gm,t)$ is the injection term.
Particles are continuously injected in the PWN during the evolution considering a broken power-law in energy, namely:
\begin{equation}\label{eq:inj}
  \Qinj(\gm,t) = Q_0(t) \times \begin{cases}
    \left(\frac{\gm}{\gmbreak}\right)^{-\alL}, & \text{if $\gm\leq \gmbreak$}\,,\\
      \\
    \left(\frac{\gm}{\gmbreak}\right)^{-\alH}, & \text{if $\gm > \gmbreak$}\,.
  \end{cases}
\end{equation}
The normalization $Q_0(t)$ is obtained at each time from the requirement that a constant fraction $(1-\eta)$ of the pulsar input goes into particles, namely:
\begin{equation}\label{eq:norm}
    (1-\eta)L(t) = \int^{\gm_{\mathrm{max}}}_{\gm_{\mathrm{min}}} \gm m_e c^2 \Qinj(\gm,t)\,d\gm\,.
\end{equation}
The minimum Lorentz factor at injection is generally assumed to equal the electron energy at rest (then $\gm_{\mathrm{min}}=1$), while the maximum Lorentz factor is computed following \citet{Martin:2016} (either considering the synchrotron acceleration limit or the maximum energy to maintain acceleration confined into the wind termination shock, see Eqs. 6-7 of that work).
The remaining fraction of the injection energy, $\eta\,L(t)$, is then converted into magnetic field at each timestep. The variation of the magnetic energy ($W_B=B^2 R^3 /6$) in the PWN is given by \citep{Martin:2016}:
\begin{equation}\label{eq_magE}
    \frac{d W_B(t)}{dt} = \eta L(t) - \frac{W_B(t)}{R(t)}\frac{d R(t)}{dt}\,.
\end{equation}
The magnetic fraction $\eta$, or better called the injection sharing,  is kept constant during the evolution, and so is the instantaneous ratio of the energy that goes into particles and into magnetic field. 
This constant injection sharing along time should not be confused with a constant energy partition, which is in fact not realizing: since particles have dynamical and radiative losses, while the magnetic energy is affected by dynamics only, the ratio of particle energy to magnetic field energy is not constant in time, nor it is equal to ($1-\eta$). 
However, TIDE provides $B(t)$, as well as the particles energy directly, and thus the energy ratio can be computed at each time (see e.g., fig. 4 of \citealt{Martin:2016}) for further discussion.
The $\eta$ parameter can be connected with the pulsar wind magnetization by: $\sigma=\eta/(1-\eta)$.

Fig.~\ref{fig:all_sp} shows the results of the spectral evolution of the selected sources. 
In the left column we show the spectrum at their actual age, the available spectral data for reference, and the relevance of the various terms contributing to the total spectrum.
In the right column we show the multi-epoch spectrum, computed at different times for the various sources, %in order 
to highlight specific moments of their reverberation process (i.e. maximum compressions or expansions).

Let us here discuss in more detail the differences between the present approach and previous ones, with particular reference to \citet{Torres_Lin:2018} (TL2018 hereafter), where the authors used TIDE (v2.2, following the nomenclature of Paper I) i.e., standard thin-shell model, as normal in literature, without lagrangian treatment of reverberation and with a different description of the shell dynamics than the one introduced in Paper I.
\begin{comment}
In that version of the code, in addition not to having a modified thin-shell evolution thanks to the semi lagrangian treatment we discussed in this work, there was a different description of the shell dynamics, as we discussed in Paper I.
\end{comment}

For the three real sources:
\begin{enumerate}
    \item \textbf{Crab nebula} (upper row). \\
        A comparison with results obtained with the standard one-zone, pure thin-shell, version of the code TIDE can be made looking at Fig.~1 in TL2018.
        The agreement is excellent at $\tage$, as we discussed previously, and the same holds at $1\,\tch$, the time at which the PWN reaches its maximum expansion (black dashed line corresponding to $t_1$ in  Fig.1 of TL2018).
        We can easily recognize in the spectrum the effects of the adiabatic expansion and magnetic field decrease in the lower synchrotron spectrum, which also cause less emission in the IC range, being the major contribution coming from self-synchrotron Compton processes (SSC).
        
        On the contrary, from the onset of reverberation, the difference with the results obtained within the pure thin-shell approximation starts to increase.
        At the maximum compression ($t_3=6841.7$ yr in TL2018, blue dashed line) the pure thin-shell modelling produces a larger compression, and thus
        on one side high energy gains 
        %due to adiabatic compression 
        (reflecting in the extra luminosity at radio frequencies in TL2018), while on the other a huge increase of the nebular magnetic field, modifying substantially the high-energy synchrotron spectrum. 
        Depending on how much the magnetic field is increased, most of the particles might be burn off, also causing a strong variation in the IC spectrum, due to the lack of electrons for the scattering events.
        On the contrary, when we release the pure thin-shell approach and model the evolution with 
        %{\bf unified} 
        TIDE+L, the maximum compression happens much later
        (at $t\sim 6\tch\sim 20$kyr, corresponding to the medium red line), 
        and we see a different scenario: the luminosity at low energies is slightly diminished while the X-ray component is almost depleted (due to aging), and the same reflects on the IC component.
        No effects connected to adiabatic gains or increase of the magnetic field can be easily identified in the full model at the spectral level. 
        From $\tage$ on, the magnetic field in the system always remains below $100\, \mu$G; the weak compressions do not cause any important field amplification.
        
        %%%%%%%%%%%%
        \item \textbf{J1834.9--0846} (middle row).
        
    \begin{figure}
	\includegraphics[width=.5\textwidth]{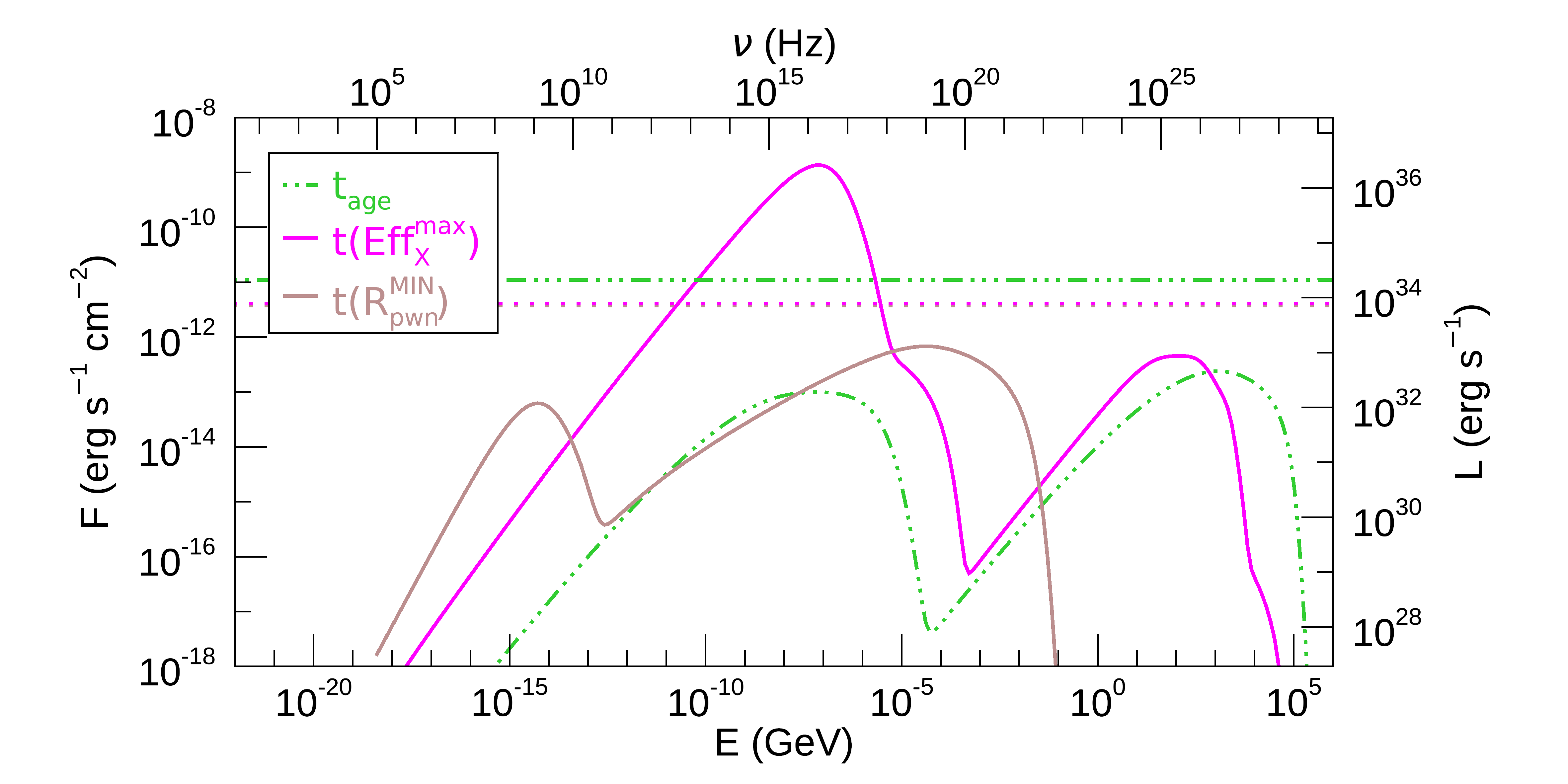}
    \caption{Plot of the spectrum of J1834.9--0846 at three different ages: $\tage$ (in dash-dotted green), $t(\mathrm{Eff}_{\mathrm{X}}^{\mathrm{max}})\simeq 2.9 \tch$, the time at which the efficiency in the X-ray band is maximum (solid magenta line) and $t(\Rpwn^{\mathrm{MIN}})$, the maximum compression (at $3.0\tch$ -- solid brown line), to be matched with the same colored line of Fig.~\ref{fig:all_sp}. Comparing with the spectral evolution of the same source discusses in TL2018, where TIDE v2.2 was used, we can notice that, despite the difference in times and intensity of the compression, the spectrum at maximum efficiency is very similar (see Fig.~1, second row, orange dashed line in that paper). The horizontal lines show the value of the spin-down luminosity of the pulsar at the three ages considered (with the same color code). The brown line, referring to $t(\Rpwn^{\mathrm{MIN}})$, is hidden in the plot by magenta line.}
    \label{fig:j18super}
\end{figure} 
        In this case, being $\tage>\tbegrev$, the variation introduced by the coupled lagrangian and thin-shell evolution of TIDE+L makes the set of spectral parameters recovered from previous works (TL2018  and Paper I) not good for a  fit to the spectral data anymore. 
        To  improve the representation, we need to substantially increase the magnetic fraction (from 0.045 to 0.15).
        Of course this also means that the comparison with previous results (same Fig.~1 in TL2018) is  no longer straightforward, and 
        not very good even at $\tage$, although the overall properties of the spectrum appear consistent.
        J1834.9--0846 is one of the two  compressible systems we considered and the one with the lowest spin-down luminosity at the maximum compression (at $t=3\tch\simeq12073$ yr).
        If we look at the spectrum at the compression peak, we recognize the effect of the adiabatic compression in the increase of the low energy emission (the bump peaking at $\sim 10^{10}$ Hz) combined with the synchrotron cooling of the higher energy synchrotron component, visible with a spectral steepening before the peak.
        If we compare with the same time from TL2018 (where the maximum of the compression happens at $t_3=8373.9$ yr), we note the same effect: 
        the extra compression induced by the pure thin-shell approximation reflects in a massive synchrotron cooling at all energies, even in the radio band (no low energy bump survives).
         However, in this case, in both approaches the compression is strong enough to increase the magnetic field to the point that most of the particles in the nebula are cooled by synchrotron losses.

        In Fig.~\ref{fig:j18super} we plot again the spectrum of J1834.9--0846 to show the detail of the maximum efficiency in X-ray emission {\color{black}(namely $L_X(t)/L(t)$)},  which the system reaches before the maximum compression (namely at a radius of $\sim 0.03\Rch$ and at $t\simeq 2.9\tch$). 
        Using the same notation as in TL2018, that time is named $t(\mathrm{Eff}_{\mathrm{X}}^{\mathrm{max}})$ and corresponds to a spectral peak at $\nu\simeq 2\times10^{16}$ Hz.
        The spin-down luminosity injected by the pulsar at that time is $L(t(\mathrm{Eff}_{\mathrm{X}}^{\mathrm{max}}))=7.8\times 10^{33}$ erg/s, while the X-ray luminosity in the 0.5-5 keV (Chandra) band is $L_X(t(\mathrm{Eff}_{\mathrm{X}}^{\mathrm{max}}))=1.6\times 10^{36}$ erg/s, namely a factor $\sim 200$ larger than the injected power, 
        %while the luminosity at the spectral peak {\bf i.e., not even integrated in any band} is $L_{\mathrm{peak}}=2.6\times 10^{36}$ erg/s, namely $333$ times larger than the pulsar injection, 
        meaning that the system will actually be going through 
        a  strong super-efficient phase.
        {\color{black}In this source, even if the magnetic field soon after reverberation is low (see Table~\ref{tab:pwnpar}), during the peak of the compression phases it is amplified to very large values, being $B\propto \Rpwn^{-2}$. In the absence of radiative losses, the first compression leads to a magnetic field of $3\,$mG, while with radiative losses it reaches  $100\,$mG. }
      %  Even if the magnetic field at the beginning of reverberation is low (see Table~\ref{tab:pwnpar}), its average value over $8 \,\tch$ is large: $\sim450 \,\mu$G ($\sim50 \,\mu$G without radiative losses). In the absence of radiative losses, only the first compression (the more intense) leads to a magnetic field larger than $100\,\mu$G (3 mG), but when radiative losses are active, all compressions lead to a huge magnetic field amplification, with the maximum value at the first compression of 0.1 G.}
    
       \item \textbf{3C58} (bottom row).
       
        At $\tbegrev\sim1.2\tch$ the spectrum shows the effects of the adiabatic expansion: the synchrotron spectrum lowers due to a combined effect of adiabatic cooling of the particles and decrease of the magnetic field.
        In this case, differently from what  we have earlier discussed for the Crab nebula, being the dominant emission at very high energies the one coming from the scattering between nebular leptons with background photons, instead of  the SSC, no variation is observed in the IC spectrum.
        At the first compression (happening at $t\sim 3.3\tch$), the spectrum shows an evident modification in morphology: the synchrotron spectrum peaks at higher frequencies due to adiabatic gains. 
        This system has been characterized a very low magnetic field, with little dynamical consequences.
         This is shown by the coincidence between the non-radiative and radiative evolution (see Fig.~\ref{fig:EVOall}). 
        It also has a rather modest compression during reverberation, and the increase in magnetic field is not enough to effectively modify 
        the IC emission  yield,
        %during compression, 
        which instead follows directly the variations of the synchrotron component.
        For 3C58 we do not have multi-epoch examples to compare with, but only the spectrum at $\tage$ as presented in \citet{Martin_Torres:2022}, with which we find a perfect agreement  for the reasons just described.    
        {\color{black}This nebula is weakly magnetized \citep{Slane:2008}, and as expected the magnetic field remains low for the entire evolution, with an average up to an age of $8\,\tch$ of $6\,\mu$G ($3\,\mu$G without losses).}
\end{enumerate}

\begin{figure*}
	\includegraphics[width=.485\textwidth]{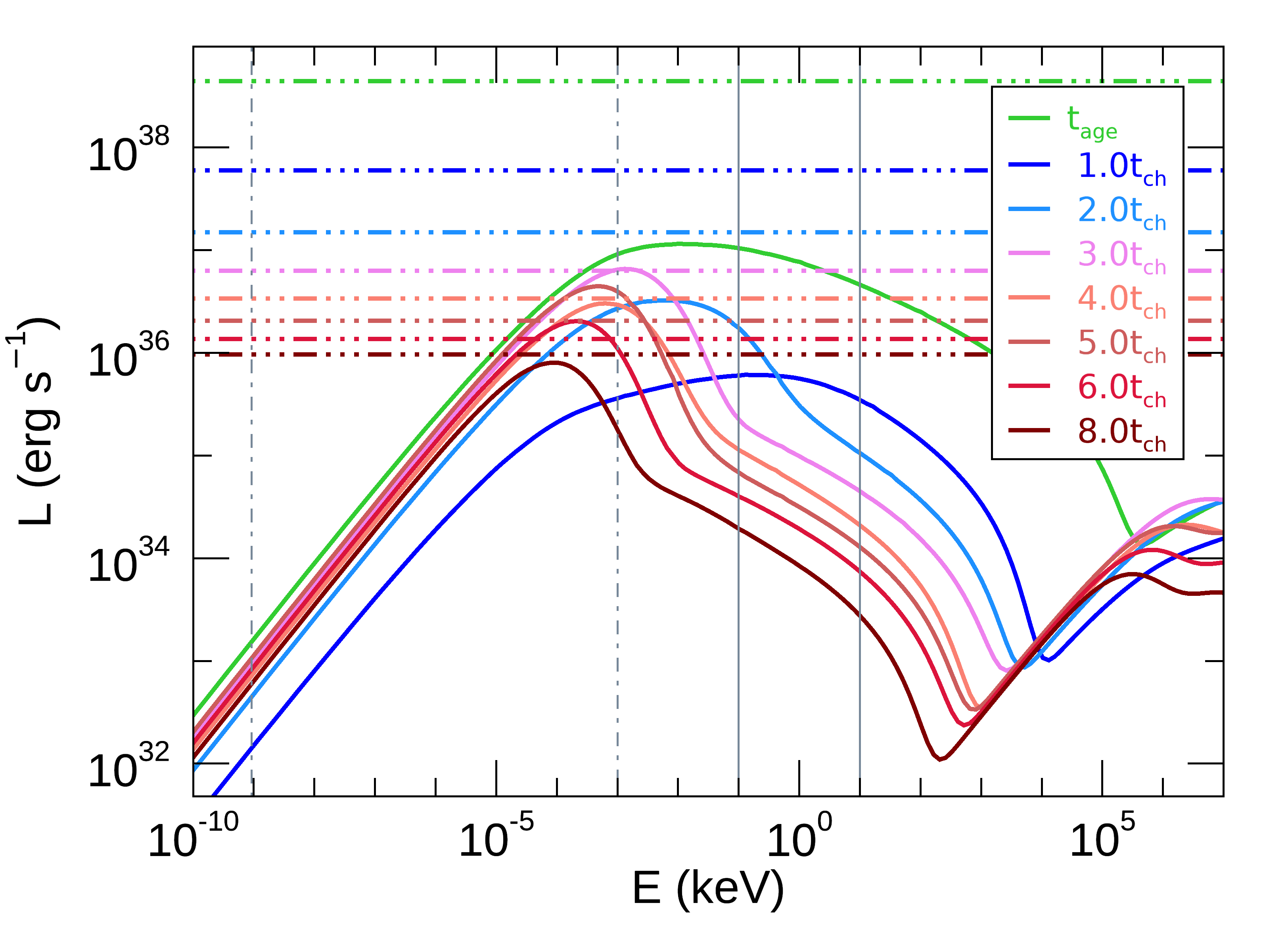}\,
 \includegraphics[width=.485\textwidth]{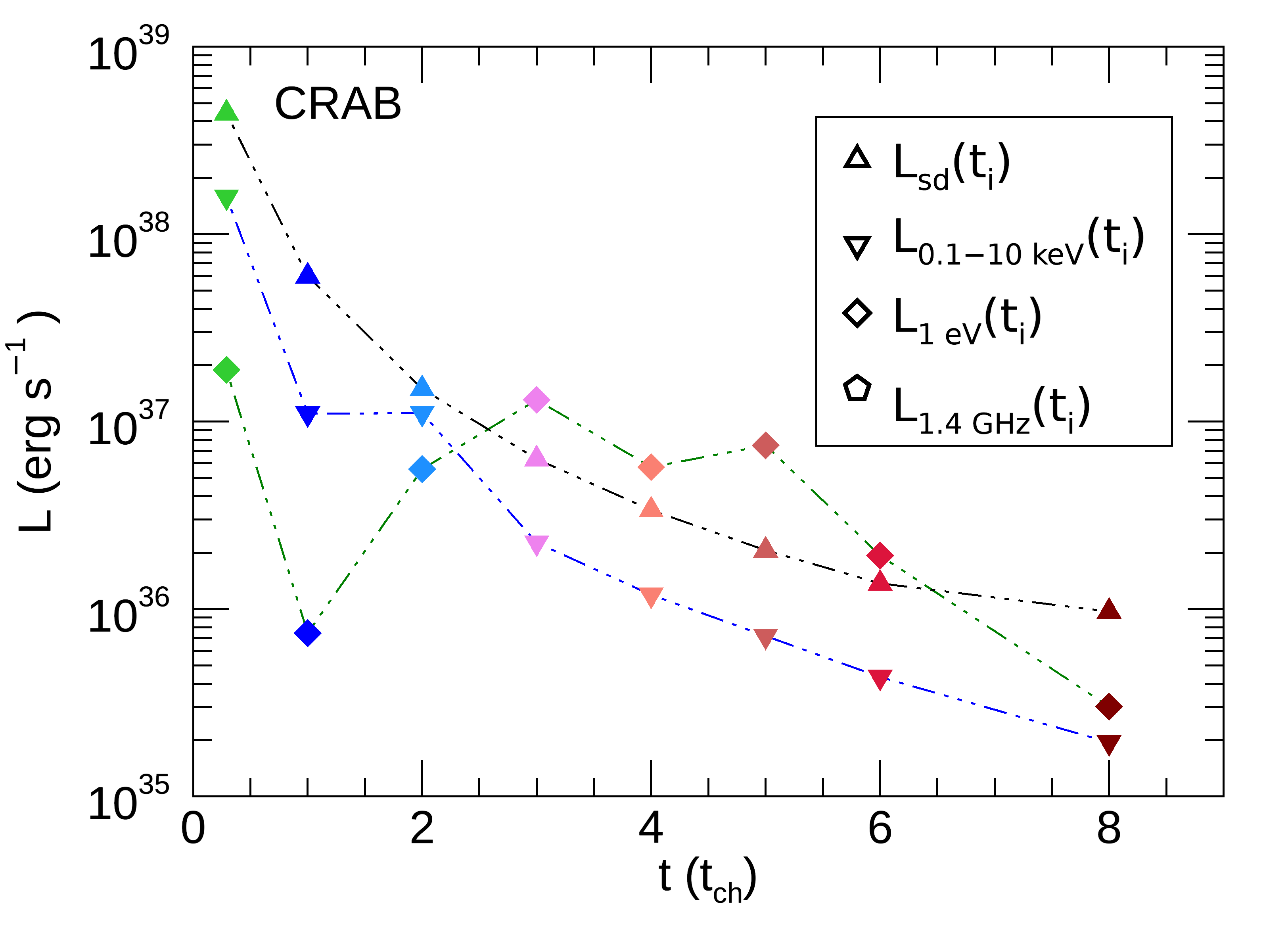}\\
 	\includegraphics[width=.485\textwidth]{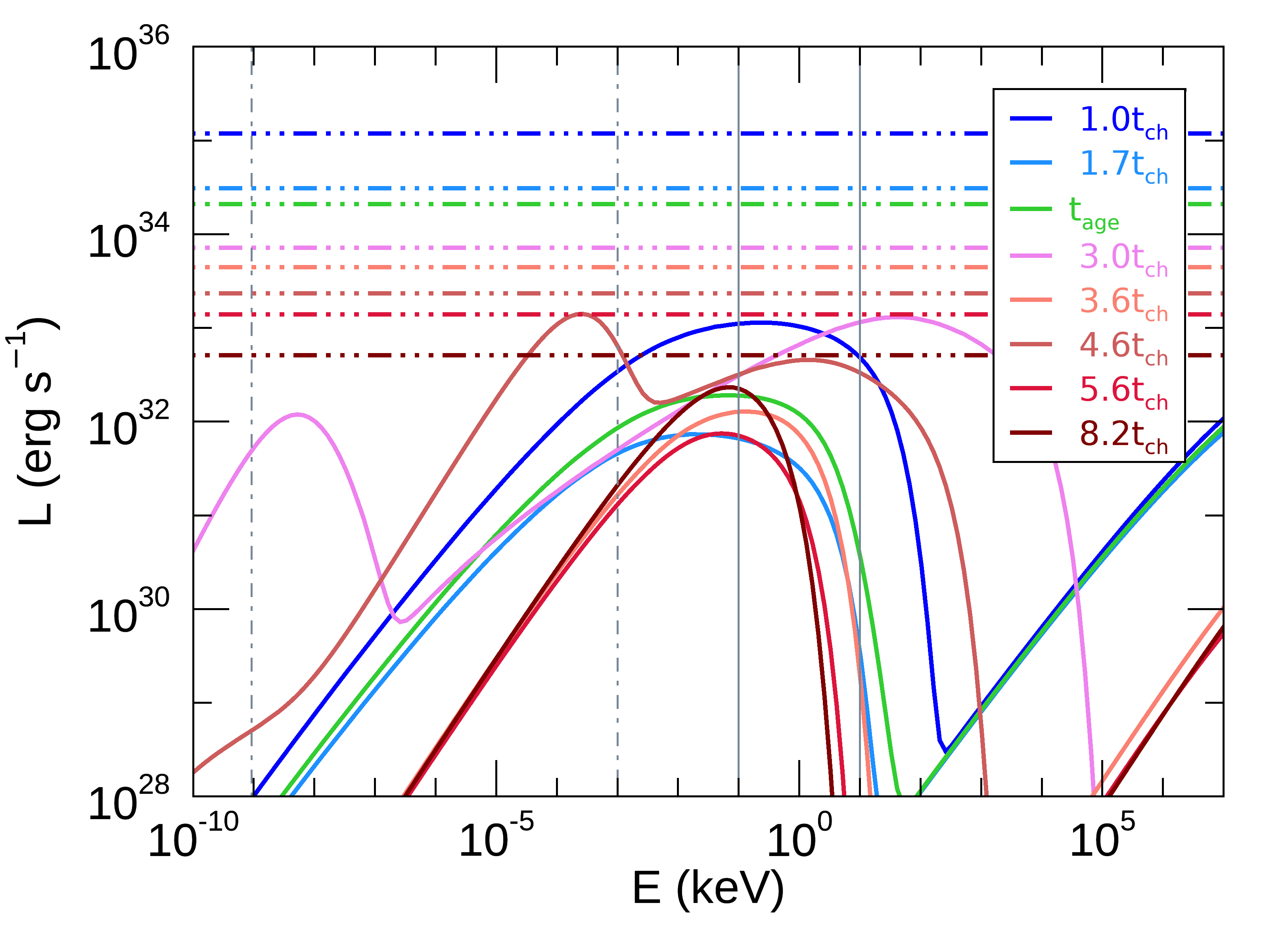}\,
 \includegraphics[width=.485\textwidth]{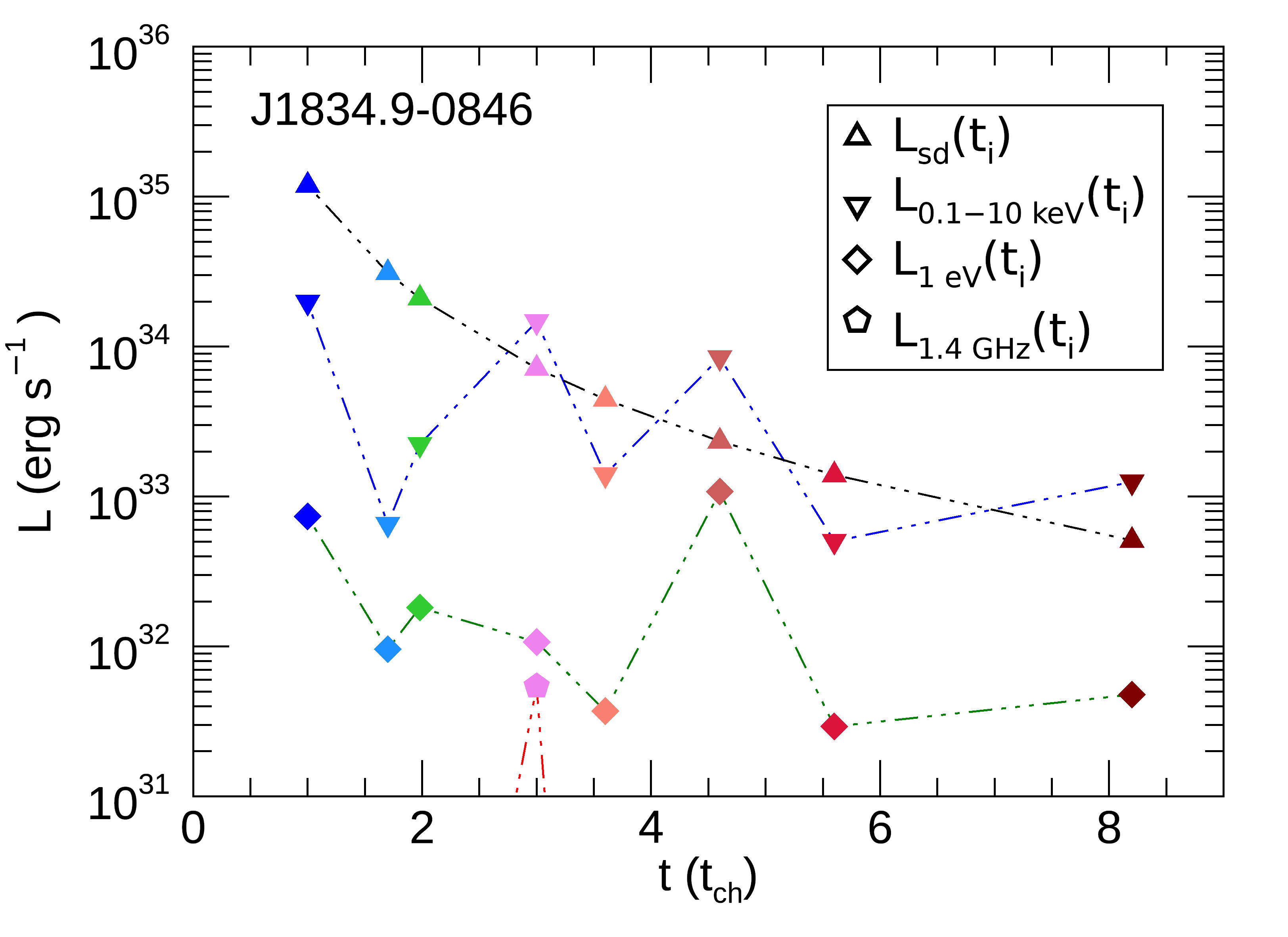}\\
  	\includegraphics[width=.485\textwidth]{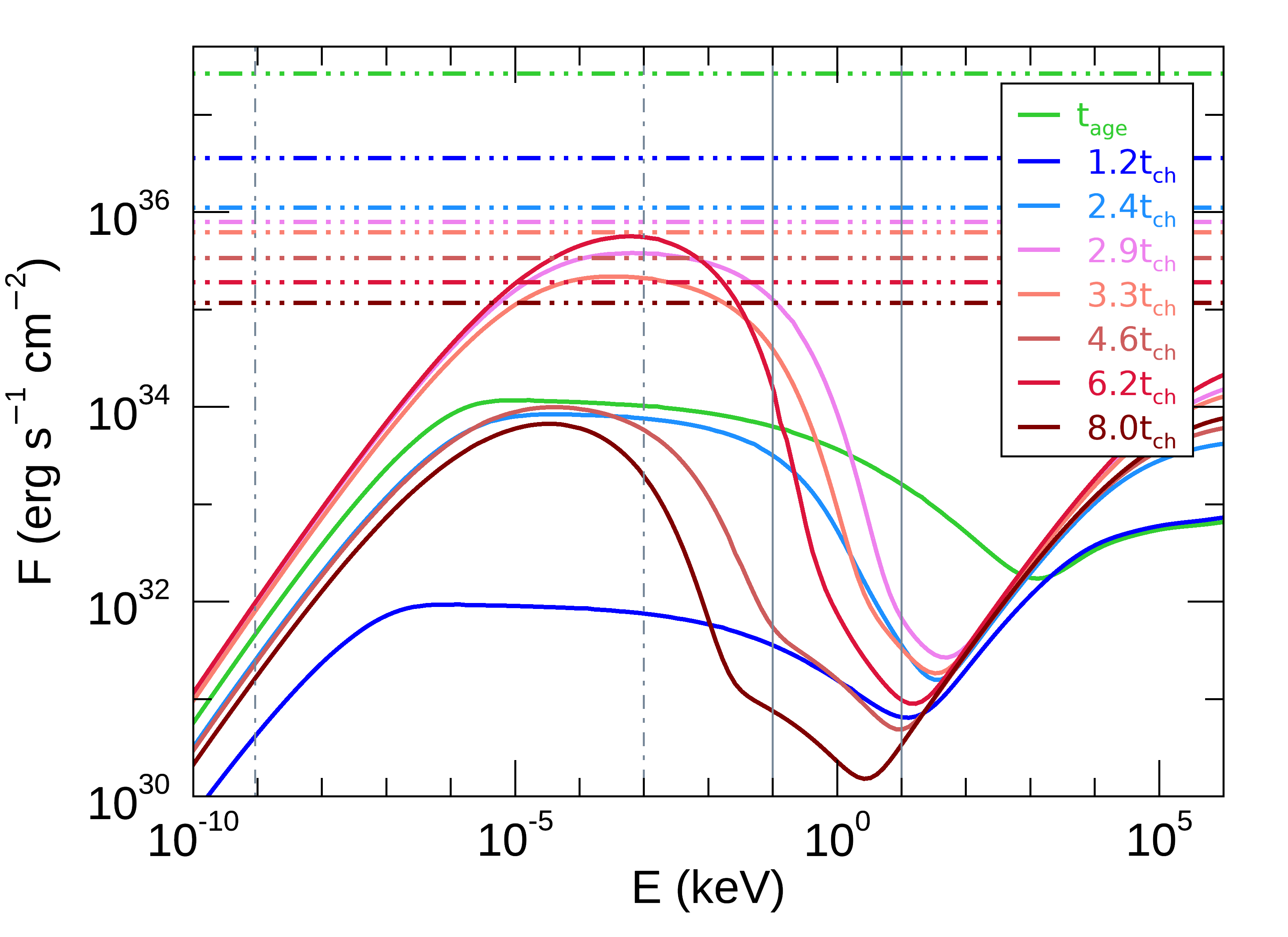}\,
 \includegraphics[width=.485\textwidth]{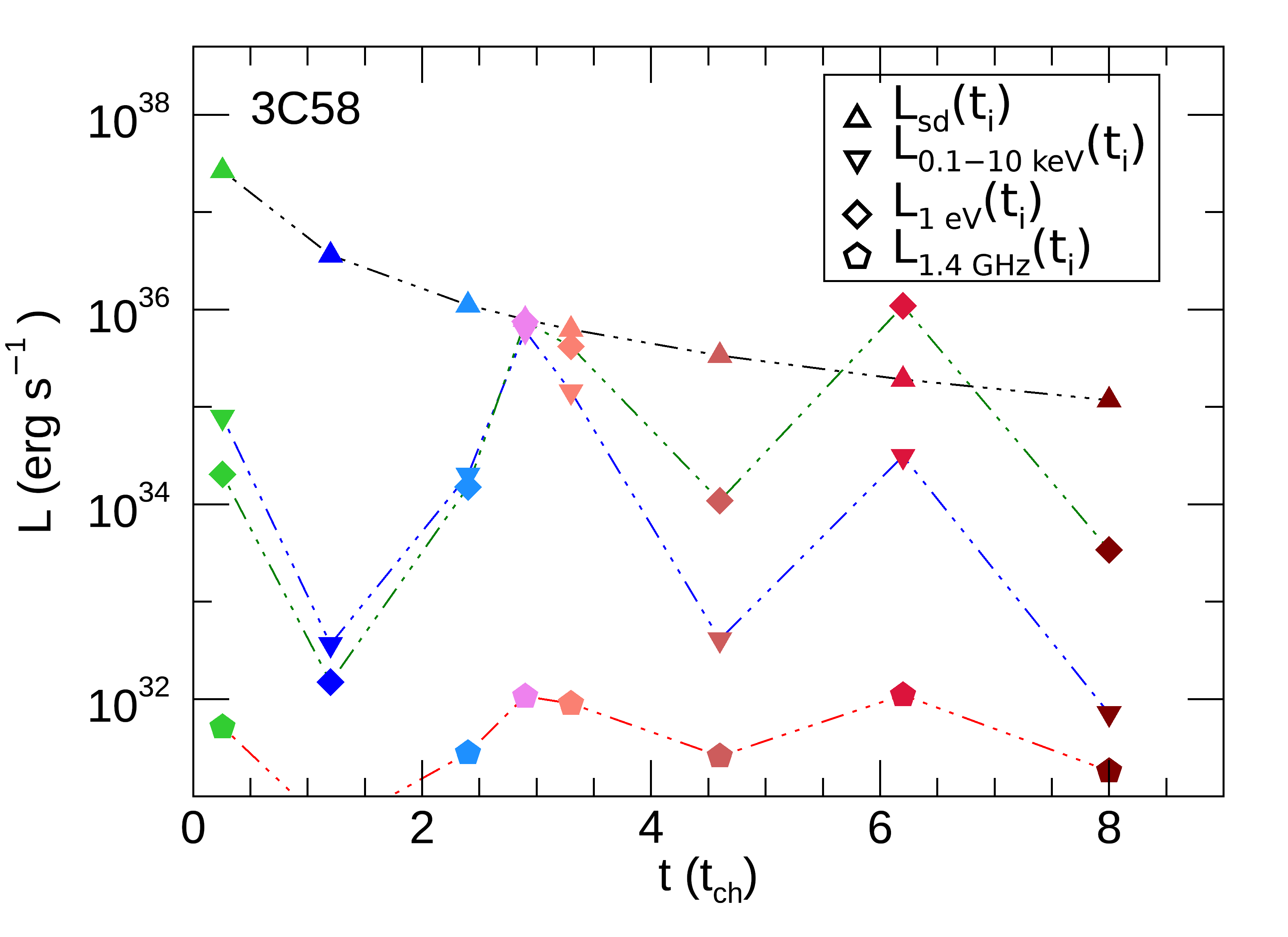}\\
    \caption{Left panels: Synchrotron spectra of the Crab (upper row), J1834.9-0846 (middle row) and 3C58 (bottom row), at the same %different 
    ages considered in  Fig.~\ref{fig:all_sp} (using the same color coding). 
    In each plot the horizontal dot-dashed lines show the value of the spin-down luminosity of the pulsar at the same age at which the spectral luminosity has been obtained (same colors for the same age).
    The vertical gray lines identify some energy bands of interest, from left to right: the radio frequency of 1.4 GHz (dot-dashed), the optical at 1 eV (dot-dashed), the 0.1-10 keV X-ray band (within the solid lines). These are the references to build up the plots of the panels on the right.
    Right panels: comparison of the integrated PWN luminosity (in radio, optical and X-rays) with the injection luminosity at the ages considered for the left plots. 
    The colors refer to the same age, while different symbols refers to different observational bands.
    Up triangles represent the pulsar spin-down at a certain age, and to facilitate reading the symbols are joined with a dot-dashed black line.
    Down triangles (joined with a dot-dashed blue line) stand for the PWN integrated X-ray luminosity, diamonds for the optical one (joined with a green dot-dashed line), pentagons for the radio luminosity (joined with a red dot-dashed line).
    Normally the spin down luminosity is larger than the PWN luminosity in any band, and up triangles are above the other symbols. 
    On the contrary, the system is facing a super-efficient phase when the symbol relative to a certain energy band stays above the up triangle (i.e. $L_i>L_{\mathrm{{sd}}}$). As it is apparent, in our case this is only happening in J1834.09-0846  at X-rays, for the 0.1-10 keV band. Interestingly, the Crab nebula shows a persistent super-efficient phase in the optical, while 3C58 has a super-efficient optical phase around $6\tch$.
    In most cases the luminosity in the radio band is much lower than other bands and it remains excluded from the plots (this is e.g. the case for the Crab, where the radio luminosity is always below $2\times 10^{33}$ erg/s), too low to extend the plot and maintain the visibility of the other points).}
    \label{fig:superff}
\end{figure*}
\noindent 
Finally, let us comment upon the possible appearance of super luminous phases in the evolution of our three sources. 
These can be easily identified in Fig.~\ref{fig:superff}, especially comparing the integrated luminosity in a given energy band with the power injected by the pulsar at the same age (panel on the right). 
As discussed previously, only J1834.9-0846 shows an  X-ray luminosity that exceeds the injection luminosity in some phases, due to the extreme compression the system undergoes in reverberation and the consequent dynamical modifications.
Interestingly, Crab and 3C58, which on the contrary do not show important modifications of their high energy emission due to the small compression, both show the presence of phases with an augmented optical emission (at 1 eV).

We believe this is an interesting feature to be further investigated in a dedicated study, extending the analysis of the spectral modifications to a larger population to identify possible observational features for future instruments working in the UV and optical-B bands.

%%%%%%%%%%%%%%%%%%%%%%%%%%%%%%%%%%%%%%%%%%%%%%%
\section{Conclusions}
\label{sec:end}
%%%%%%%%%%%%%%%%%%%%%%%%%%%%%%%%%%%%%%%%%%%%%%%

Despite the fact that middle-aged PWNe constitute the vast majority of all PWN-SNR systems, and the fact that perhaps they form one of the main contributor to the diffuse Galactic gamma-ray background, little is known of their evolutionary properties. This is in sharp contrast with younger systems for which, thanks to a plethora of approaches, we have now a well established canonical picture.

Barring a few limited numerical multi-dimensional simulations of selected systems, currently the most popular approach to investigate the late time evolution of PWNe is within the so called thin-shell approximation. However a correct description of the dynamics of the compression  that a PWN experiences during reverberation
requires a correct description of both the shell properties and that of the SNR surrounding it. In Paper II, in fact, we showed how sensitive the PWN evolution is to a rough representation of both {\it of the latter.}
This not only affects the shape and size of the PWN, but also the future evolution of the relativistic pair population, and thus the spectral properties of such systems, together with radiative losses which feedback on the dynamical bouncing properties.

In the present work we have further extended our study of the properties of post-reverberation systems, 
introducing a new  numerical technique for the dynamical-spectral evolution of a PWN interacting with its SNR through and beyond the reverberation phase.
Using this hybrid approach, that combines lagrangian techniques for the dynamical evolution of the SNR, with standard thin-shell evolution for the PWN, together with time dependent radiative model for the spectral properties of the PWN, we are able to model self-consistently the entire  
evolution, allowing PWNe simulations to extend up to the typical age range where most of the TeV population is found.

The code presented here is based on a modified version of TIDE, and its performances, with respect to a more sophisticated full-lagrangian approach, has been cross-checked for a variety of input conditions that cover the parameters space of the PWN-SNR population. The new approach correctly recovers the full evolution both during free-expansion and later reverberation.

As it might be expected, we found that radiative losses introduce complexity to the description of the PWN response to the SNR in reverberation, especially for highly-compressible systems: sources at different position in the PWN-SNR parameters space can show similar compression factors, due to the combination of their injection parameters ($L_0, \tau_0$), the SNR characteristic age ($\tch$ and how it compares with $\tau_0$) and the relevance of synchrotron losses (the magnetization parameter $\eta$). 
A general treatment of the effect of radiation losses and its possible parametrization require a much more detailed investigation of the parameters space, that goes beyond the scope of the present paper and that we leave for future investigations.

The importance of the results presented here is immediately clear when we look at the multi-band spectral evolution of the sources, in the later evolutionary phases. While in the free-expansion results are unchanged, substantial differences arises at later times, especially for the low compressive systems. The difference is less critical for highly-compressible cases. For these cases, we confirm the existence of the so called super-efficient phases.

The minor compression reflects in a less structured spectrum of the sources at the different ages, since most of the bumps and variations to the spectral slopes observed with previous models arise from the excessive compression, causing increased synchrotron cooling and appearance of bumps connected with the adiabatic gains in the strong compression phases.
We plan to use our new model to produce reliable PWNe population studies, extending the concept presented in \citet{Fiori:2021} to multi-wavelengths analysis,  as well as gather estimates of the extent and the number of super-efficient PWNe existing in the Galaxy, enhancing the work by \cite{Torres2019}.

%%%%%%%%%%%%%%%%%%%%%%%%%%%%%%%%%%%%%%%%%%%%
\section*{Acknowledgements}
%%%%%%%%%%%%%%%%%%%%%%%%%%%%%%%%%%%%%%%%%%%%
%
%
This work has been supported by INAF grants: MAINSTREAM 2018 \textit{Particle Acceleration in Galactic Sources in the CTA era}, MiniGrant PWNnumpol - \textit{Numerical Studies of Pulsar Wind Nebulae in The Light of IXPE}, PRIN-INAF 2019 \textit{From massive stars to supernovae and supernova remnants: driving mass, energy and cosmic rays in our Galaxy}, and by Spanish grants PID2021-124581OB-I00 funded by MCIN/AEI/10.13039/501100011033, 2021SGR00426 of the Generalitat de Catalunya, by the program Unidad de Excelencia María de Maeztu CEX2020-001058-M, and by 
MCIN with funding from European Union NextGeneration EU (PRTR-C17.I1).
%

%%%%%%%%%%%%%%%%%%%%%%%%%%%%%%%%%%%%%%%%%%%%%%%%%%

\section*{Data availability}
The data for the models underlying this article will be shared on reasonable request to the corresponding authors.

% References
%%%%%%%%%%%%%%%%%%%%%%%%%%%%%%%%%%%%%%%%%%%%%%%%%%

\bibliographystyle{mn2e} 
\bibliography{biblio}

\label{lastpage}
%%%%%%%%%%%%%%%%%%%%%%%%%%%%%%%%%%%%%%%%%%%%%%%%
\end{document}